\documentclass[longauth]{aa}  

\usepackage{lipsum}
\usepackage{graphicx}
\usepackage{txfonts}
\usepackage{xspace}
\usepackage{enumitem}
\usepackage[colorlinks=true,allcolors=blue]{hyperref}
\begin{document}

   \title{The CHEOPS view of HD 95338b: refined transit parameters, and a search for exomoons\thanks{Based on data from CHEOPS Guaranteed Time Observations, collected under Programme ID CH\textunderscore PR100009 and CH\textunderscore PR140072. The raw and detrended photometric time series data are available in electronic form at the CDS via anonymous ftp to ???????}}

\author{
Sz.~K\'alm\'an\inst{\ref{inst:13}, \ref{inst:14}, \ref{inst:6}, \ref{inst:phd}}\,$^{\href{https://orcid.org/0000-0003-3754-7889}{\protect\includegraphics[height=0.19cm]{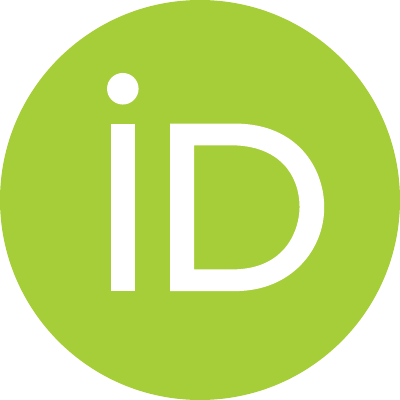}}}$\and 
A.~E.~Simon\inst{\ref{inst:1},\ref{inst:2}}\,$^{\href{https://orcid.org/0000-0001-9773-2600}{\protect\includegraphics[height=0.19cm]{figures/orcid.pdf}}}$\and 
A.~Deline\inst{\ref{inst:3}}\and 
Sz.~Csizmadia\inst{\ref{inst:4}}\,$^{\href{https://orcid.org/0000-0001-6803-9698}{\protect\includegraphics[height=0.19cm]{figures/orcid.pdf}}}$\and 
Gy.~M.~Szabó\inst{\ref{inst:5},\ref{inst:6}}\,$^{\href{https://orcid.org/0000-0002-0606-7930}{\protect\includegraphics[height=0.19cm]{figures/orcid.pdf}}}$\and 
D.~Ehrenreich\inst{\ref{inst:3},\ref{inst:7}}\,$^{\href{https://orcid.org/0000-0001-9704-5405}{\protect\includegraphics[height=0.19cm]{figures/orcid.pdf}}}$\and 
T.~G.~Wilson\inst{\ref{inst:8}}\,$^{\href{https://orcid.org/0000-0001-8749-1962}{\protect\includegraphics[height=0.19cm]{figures/orcid.pdf}}}$\and 
M.~N.~Günther\inst{\ref{inst:9}}\,$^{\href{https://orcid.org/0000-0002-3164-9086}{\protect\includegraphics[height=0.19cm]{figures/orcid.pdf}}}$\and 
A.~Heitzmann\inst{\ref{inst:3}}\,$^{\href{https://orcid.org/0000-0002-8091-7526}{\protect\includegraphics[height=0.19cm]{figures/orcid.pdf}}}$\and 
S.~G.~Sousa\inst{\ref{inst:10}}\,$^{\href{https://orcid.org/0000-0001-9047-2965}{\protect\includegraphics[height=0.19cm]{figures/orcid.pdf}}}$\and 
M.~Farnir\inst{\ref{inst:11}}\and 
A.~Bonfanti\inst{\ref{inst:12}}\,$^{\href{https://orcid.org/0000-0002-1916-5935}{\protect\includegraphics[height=0.19cm]{figures/orcid.pdf}}}$\and 
A.~M.~S.~Smith\inst{\ref{inst:4}}\,$^{\href{https://orcid.org/0000-0002-2386-4341}{\protect\includegraphics[height=0.19cm]{figures/orcid.pdf}}}$\and 
A.~Pal\inst{\ref{inst:13},\ref{inst:14},\ref{inst:6}}\and 
G.~Scandariato\inst{\ref{inst:15}}\,$^{\href{https://orcid.org/0000-0003-2029-0626}{\protect\includegraphics[height=0.19cm]{figures/orcid.pdf}}}$\and 
V.~Adibekyan\inst{\ref{inst:10}}\,$^{\href{https://orcid.org/0000-0002-0601-6199}{\protect\includegraphics[height=0.19cm]{figures/orcid.pdf}}}$\and 
A.~Brandeker\inst{\ref{inst:16}}\,$^{\href{https://orcid.org/0000-0002-7201-7536}{\protect\includegraphics[height=0.19cm]{figures/orcid.pdf}}}$\and 
S.~Charnoz\inst{\ref{inst:charn}} \and
B.~Akinsanmi\inst{\ref{inst:3}}\,$^{\href{https://orcid.org/0000-0001-6519-1598}{\protect\includegraphics[height=0.19cm]{figures/orcid.pdf}}}$\and 
S.~C.~C.~Barros\inst{\ref{inst:10},\ref{inst:21}}\,$^{\href{https://orcid.org/0000-0003-2434-3625}{\protect\includegraphics[height=0.19cm]{figures/orcid.pdf}}}$\and 
X. Song\inst{\ref{inst:3}, \ref{inst:beijing}}\,$^{\href{https://orcid.org/0000-0003-4972-7772}{\protect \includegraphics[height = 0.19 cm]{figures/orcid.pdf}}}$\and
Y.~Alibert\inst{\ref{inst:2},\ref{inst:1}}\,$^{\href{https://orcid.org/0000-0002-4644-8818}{\protect\includegraphics[height=0.19cm]{figures/orcid.pdf}}}$\and 
R.~Alonso\inst{\ref{inst:17},\ref{inst:18}}\,$^{\href{https://orcid.org/0000-0001-8462-8126}{\protect\includegraphics[height=0.19cm]{figures/orcid.pdf}}}$\and 
T.~Bárczy\inst{\ref{inst:19}}\,$^{\href{https://orcid.org/0000-0002-7822-4413}{\protect\includegraphics[height=0.19cm]{figures/orcid.pdf}}}$\and 
D.~Barrado~Navascues\inst{\ref{inst:20}}\,$^{\href{https://orcid.org/0000-0002-5971-9242}{\protect\includegraphics[height=0.19cm]{figures/orcid.pdf}}}$\and 
W.~Baumjohann\inst{\ref{inst:12}}\,$^{\href{https://orcid.org/0000-0001-6271-0110}{\protect\includegraphics[height=0.19cm]{figures/orcid.pdf}}}$\and 
W.~Benz\inst{\ref{inst:1},\ref{inst:2}}\,$^{\href{https://orcid.org/0000-0001-7896-6479}{\protect\includegraphics[height=0.19cm]{figures/orcid.pdf}}}$\and 
N.~Billot\inst{\ref{inst:3}}\,$^{\href{https://orcid.org/0000-0003-3429-3836}{\protect\includegraphics[height=0.19cm]{figures/orcid.pdf}}}$\and 
F.~Biondi\inst{\ref{inst:mpla},\ref{inst:22}}\,$^{\href{https://orcid.org/0000-0002-1337-3653}{\protect\includegraphics[height=0.19cm]{figures/orcid.pdf}}}$\and 
L.~Borsato\inst{\ref{inst:22}}\,$^{\href{https://orcid.org/0000-0003-0066-9268}{\protect\includegraphics[height=0.19cm]{figures/orcid.pdf}}}$\and 
C.~Broeg\inst{\ref{inst:1},\ref{inst:2}}\,$^{\href{https://orcid.org/0000-0001-5132-2614}{\protect\includegraphics[height=0.19cm]{figures/orcid.pdf}}}$\and 
A.~Collier~Cameron\inst{\ref{inst:23}}\,$^{\href{https://orcid.org/0000-0002-8863-7828}{\protect\includegraphics[height=0.19cm]{figures/orcid.pdf}}}$\and 
C.~Corral~van~Damme\inst{\ref{inst:45}}\and
A.~C.~M.~Correia\inst{\ref{inst:24}}\,$^{\href{https://orcid.org/0000-0002-8946-8579}{\protect\includegraphics[height=0.19cm]{figures/orcid.pdf}}}$\and 
P.~E.~Cubillos\inst{\ref{inst:12},\ref{inst:25}}\and 
M.~B.~Davies\inst{\ref{inst:26}}\,$^{\href{https://orcid.org/0000-0001-6080-1190}{\protect\includegraphics[height=0.19cm]{figures/orcid.pdf}}}$\and 
M.~Deleuil\inst{\ref{inst:27}}\,$^{\href{https://orcid.org/0000-0001-6036-0225}{\protect\includegraphics[height=0.19cm]{figures/orcid.pdf}}}$\and 
O.~D.~S.~Demangeon\inst{\ref{inst:10},\ref{inst:21}}\,$^{\href{https://orcid.org/0000-0001-7918-0355}{\protect\includegraphics[height=0.19cm]{figures/orcid.pdf}}}$\and 
B.-O.~Demory\inst{\ref{inst:2},\ref{inst:1}}\,$^{\href{https://orcid.org/0000-0002-9355-5165}{\protect\includegraphics[height=0.19cm]{figures/orcid.pdf}}}$\and 
A.~Derekas\inst{\ref{inst:5}}\and 
B.~Edwards\inst{\ref{inst:28}}\and 
J.~A.~Egger\inst{\ref{inst:1}}\,$^{\href{https://orcid.org/0000-0003-1628-4231}{\protect\includegraphics[height=0.19cm]{figures/orcid.pdf}}}$\and 
A.~Erikson\inst{\ref{inst:4}}\and 
A.~Fortier\inst{\ref{inst:1},\ref{inst:2}}\,$^{\href{https://orcid.org/0000-0001-8450-3374}{\protect\includegraphics[height=0.19cm]{figures/orcid.pdf}}}$\and 
L.~Fossati\inst{\ref{inst:12}}\,$^{\href{https://orcid.org/0000-0003-4426-9530}{\protect\includegraphics[height=0.19cm]{figures/orcid.pdf}}}$\and 
M.~Fridlund\inst{\ref{inst:31},\ref{inst:32}}\,$^{\href{https://orcid.org/0000-0002-0855-8426}{\protect\includegraphics[height=0.19cm]{figures/orcid.pdf}}}$\and 
D.~Gandolfi\inst{\ref{inst:33}}\,$^{\href{https://orcid.org/0000-0001-8627-9628}{\protect\includegraphics[height=0.19cm]{figures/orcid.pdf}}}$\and 
K.~Gazeas\inst{\ref{inst:34}}\,$^{\href{https://orcid.org/0000-0002-8855-3923}{\protect\includegraphics[height=0.19cm]{figures/orcid.pdf}}}$\and 
M.~Gillon\inst{\ref{inst:35}}\,$^{\href{https://orcid.org/0000-0003-1462-7739}{\protect\includegraphics[height=0.19cm]{figures/orcid.pdf}}}$\and 
M.~Güdel\inst{\ref{inst:36}}\and 
P.~Guterman\inst{\ref{inst:27},\ref{inst:37}}\and 
J.~Hasiba\inst{\ref{inst:12}}\and 
Ch.~Helling\inst{\ref{inst:12},\ref{inst:38}}\and 
K.~G.~Isaak\inst{\ref{inst:9}}\,$^{\href{https://orcid.org/0000-0001-8585-1717}{\protect\includegraphics[height=0.19cm]{figures/orcid.pdf}}}$\and 
L.~L.~Kiss\inst{\ref{inst:13},\ref{inst:14},\ref{inst:40}}\and 
J.~Korth\inst{\ref{inst:3}}\,$^{\href{https://orcid.org/0000-0002-0076-6239}{\protect\includegraphics[height=0.19cm]{figures/orcid.pdf}}}$\and 
K.~W.~F.~Lam\inst{\ref{inst:4}}\,$^{\href{https://orcid.org/0000-0002-9910-6088}{\protect\includegraphics[height=0.19cm]{figures/orcid.pdf}}}$\and 
J.~Laskar\inst{\ref{inst:42}}\,$^{\href{https://orcid.org/0000-0003-2634-789X}{\protect\includegraphics[height=0.19cm]{figures/orcid.pdf}}}$\and 
A.~Lecavelier~des~Etangs\inst{\ref{inst:43}}\,$^{\href{https://orcid.org/0000-0002-5637-5253}{\protect\includegraphics[height=0.19cm]{figures/orcid.pdf}}}$\and 
A.~Leleu\inst{\ref{inst:3},\ref{inst:1}}\,$^{\href{https://orcid.org/0000-0003-2051-7974}{\protect\includegraphics[height=0.19cm]{figures/orcid.pdf}}}$\and 
M.~Lendl\inst{\ref{inst:3}}\,$^{\href{https://orcid.org/0000-0001-9699-1459}{\protect\includegraphics[height=0.19cm]{figures/orcid.pdf}}}$\and 
D.~Magrin\inst{\ref{inst:22}}\,$^{\href{https://orcid.org/0000-0003-0312-313X}{\protect\includegraphics[height=0.19cm]{figures/orcid.pdf}}}$\and 
P.~F.~L.~Maxted\inst{\ref{inst:44}}\,$^{\href{https://orcid.org/0000-0003-3794-1317}{\protect\includegraphics[height=0.19cm]{figures/orcid.pdf}}}$\and 
B.~Merín\inst{\ref{inst:45}}\,$^{\href{https://orcid.org/0000-0002-8555-3012}{\protect\includegraphics[height=0.19cm]{figures/orcid.pdf}}}$\and 
C.~Mordasini\inst{\ref{inst:1},\ref{inst:2}}\and 
M.~Munari\inst{\ref{inst:15}}\,$^{\href{https://orcid.org/0000-0003-0990-050X}{\protect\includegraphics[height=0.19cm]{figures/orcid.pdf}}}$\and 
V.~Nascimbeni\inst{\ref{inst:22}}\,$^{\href{https://orcid.org/0000-0001-9770-1214}{\protect\includegraphics[height=0.19cm]{figures/orcid.pdf}}}$\and 
G.~Olofsson\inst{\ref{inst:16}}\,$^{\href{https://orcid.org/0000-0003-3747-7120}{\protect\includegraphics[height=0.19cm]{figures/orcid.pdf}}}$\and 
R.~Ottensamer\inst{\ref{inst:36}}\and 
I.~Pagano\inst{\ref{inst:15}}\,$^{\href{https://orcid.org/0000-0001-9573-4928}{\protect\includegraphics[height=0.19cm]{figures/orcid.pdf}}}$\and 
E.~Pallé\inst{\ref{inst:17},\ref{inst:18}}\,$^{\href{https://orcid.org/0000-0003-0987-1593}{\protect\includegraphics[height=0.19cm]{figures/orcid.pdf}}}$\and 
G.~Peter\inst{\ref{inst:4}}\,$^{\href{https://orcid.org/0000-0001-6101-2513}{\protect\includegraphics[height=0.19cm]{figures/orcid.pdf}}}$\and 
D.~Piazza\inst{\ref{inst:47}}\and 
G.~Piotto\inst{\ref{inst:22},\ref{inst:48}}\,$^{\href{https://orcid.org/0000-0002-9937-6387}{\protect\includegraphics[height=0.19cm]{figures/orcid.pdf}}}$\and 
D.~Pollacco\inst{\ref{inst:8}}\and 
D.~Queloz\inst{\ref{inst:49},\ref{inst:50}}\,$^{\href{https://orcid.org/0000-0002-3012-0316}{\protect\includegraphics[height=0.19cm]{figures/orcid.pdf}}}$\and 
R.~Ragazzoni\inst{\ref{inst:22},\ref{inst:48}}\,$^{\href{https://orcid.org/0000-0002-7697-5555}{\protect\includegraphics[height=0.19cm]{figures/orcid.pdf}}}$\and 
N.~Rando\inst{\ref{inst:9}}\and 
H.~Rauer\inst{\ref{inst:dlr_main},\ref{inst:51}}\,$^{\href{https://orcid.org/0000-0002-6510-1828}{\protect\includegraphics[height=0.19cm]{figures/orcid.pdf}}}$\and 
I.~Ribas\inst{\ref{inst:52},\ref{inst:53}}\,$^{\href{https://orcid.org/0000-0002-6689-0312}{\protect\includegraphics[height=0.19cm]{figures/orcid.pdf}}}$\and 
N.~C.~Santos\inst{\ref{inst:10},\ref{inst:21}}\,$^{\href{https://orcid.org/0000-0003-4422-2919}{\protect\includegraphics[height=0.19cm]{figures/orcid.pdf}}}$\and 
D.~Ségransan\inst{\ref{inst:3}}\,$^{\href{https://orcid.org/0000-0003-2355-8034}{\protect\includegraphics[height=0.19cm]{figures/orcid.pdf}}}$\and 
M.~Stalport\inst{\ref{inst:11},\ref{inst:35}}\and 
S.~Sulis\inst{\ref{inst:27}}\,$^{\href{https://orcid.org/0000-0001-8783-526X}{\protect\includegraphics[height=0.19cm]{figures/orcid.pdf}}}$\and 
S.~Udry\inst{\ref{inst:3}}\,$^{\href{https://orcid.org/0000-0001-7576-6236}{\protect\includegraphics[height=0.19cm]{figures/orcid.pdf}}}$\and
B.~Ulmer\inst{\ref{inst:47}}\and 
S.~Ulmer-Moll\inst{\ref{inst:3},\ref{inst:1},\ref{inst:11}}\,$^{\href{https://orcid.org/0000-0003-2417-7006}{\protect\includegraphics[height=0.19cm]{figures/orcid.pdf}}}$\and 
V.~Van~Grootel\inst{\ref{inst:11}}\,$^{\href{https://orcid.org/0000-0003-2144-4316}{\protect\includegraphics[height=0.19cm]{figures/orcid.pdf}}}$\and 
J.~Venturini\inst{\ref{inst:3}}\,$^{\href{https://orcid.org/0000-0001-9527-2903}{\protect\includegraphics[height=0.19cm]{figures/orcid.pdf}}}$\and 
E.~Villaver\inst{\ref{inst:17},\ref{inst:18}}\and 
N.~A.~Walton\inst{\ref{inst:54}}\,$^{\href{https://orcid.org/0000-0003-3983-8778}{\protect\includegraphics[height=0.19cm]{figures/orcid.pdf}}}$\and 
S.~Wolf\inst{\ref{inst:47}}\and
T.~Zingales\inst{\ref{inst:48},\ref{inst:22}}\,$^{\href{https://orcid.org/0000-0001-6880-5356}{\protect\includegraphics[height=0.19cm]{figures/orcid.pdf}}}$
}

\institute{
\label{inst:13} Konkoly Observatory, HUN-REN Research Centre for Astronomy and Earth Sciences, Konkoly Thege út 15-17., H-1121, Budapest, Hungary \and
\label{inst:14} CSFK, MTA Centre of Excellence, Budapest, Konkoly Thege út 15-17., H-1121, Hungary \and
\label{inst:6} HUN-REN-ELTE Exoplanet Research Group, Szent Imre h. u. 112., Szombathely, H-9700, Hungary \and
\label{inst:phd} ELTE E\"otv\"os Lor\'and University, Doctoral School of Physics, Budapest, H-1117 Budapest P\'azm\'any P\'eter s\'et\'any 1/A, Hungary
\label{inst:1} Space Research and Planetary Sciences, Physics Institute, University of Bern, Gesellschaftsstrasse 6, 3012 Bern, Switzerland \and
\label{inst:2} Center for Space and Habitability, University of Bern, Gesellschaftsstrasse 6, 3012 Bern, Switzerland \and
\label{inst:3} Observatoire astronomique de l'Université de Genève, Chemin Pegasi 51, 1290 Versoix, Switzerland \and
\label{inst:4} Institute of Space Research, German Aerospace Center (DLR), Rutherfordstrasse 2, 12489 Berlin, Germany \and
\label{inst:5} ELTE Gothard Astrophysical Observatory, 9700 Szombathely, Szent Imre h. u. 112, Hungary \and
\label{inst:7} Centre Vie dans l’Univers, Faculté des sciences, Université de Genève, Quai Ernest-Ansermet 30, 1211 Genève 4, Switzerland \and
\label{inst:8} Department of Physics, University of Warwick, Gibbet Hill Road, Coventry CV4 7AL, United Kingdom \and
\label{inst:9} European Space Agency (ESA), European Space Research and Technology Centre (ESTEC), Keplerlaan 1, 2201 AZ Noordwijk, The Netherlands \and
\label{inst:10} Instituto de Astrofisica e Ciencias do Espaco, Universidade do Porto, CAUP, Rua das Estrelas, 4150-762 Porto, Portugal \and
\label{inst:11} Space sciences, Technologies and Astrophysics Research (STAR) Institute, Université de Liège, Allée du 6 Août 19C, 4000 Liège, Belgium \and
\label{inst:12} Space Research Institute, Austrian Academy of Sciences, Schmiedlstrasse 6, A-8042 Graz, Austria \and
\label{inst:15} INAF, Osservatorio Astrofisico di Catania, Via S. Sofia 78, 95123 Catania, Italy \and
\label{inst:16} Department of Astronomy, Stockholm University, AlbaNova University Center, 10691 Stockholm, Sweden \and
\label{inst:17} Instituto de Astrofísica de Canarias, Vía Láctea s/n, 38200 La Laguna, Tenerife, Spain \and
\label{inst:18} Departamento de Astrofísica, Universidad de La Laguna, Astrofísico Francisco Sanchez s/n, 38206 La Laguna, Tenerife, Spain \and
\label{inst:19} Admatis, 5. Kandó Kálmán Street, 3534 Miskolc, Hungary \and
\label{inst:20} Depto. de Astrofísica, Centro de Astrobiología (CSIC-INTA), ESAC campus, 28692 Villanueva de la Cañada (Madrid), Spain \and
\label{inst:21} Departamento de Fisica e Astronomia, Faculdade de Ciencias, Universidade do Porto, Rua do Campo Alegre, 4169-007 Porto, Portugal \and
\label{inst:mpla} Max Planck Institute for Extraterrestrial Physics,
Gießenbachstraße 1, Garching bei München 85748, Germany \and
\label{inst:22} INAF, Osservatorio Astronomico di Padova, Vicolo dell'Osservatorio 5, 35122 Padova, Italy \and
\label{inst:23} Centre for Exoplanet Science, SUPA School of Physics and Astronomy, University of St Andrews, North Haugh, St Andrews KY16 9SS, UK \and
\label{inst:24} CFisUC, Departamento de Física, Universidade de Coimbra, 3004-516 Coimbra, Portugal \and
\label{inst:25} INAF, Osservatorio Astrofisico di Torino, Via Osservatorio, 20, I-10025 Pino Torinese To, Italy \and
\label{inst:26} Centre for Mathematical Sciences, Lund University, Box 118, 221 00 Lund, Sweden \and
\label{inst:27} Aix Marseille Univ, CNRS, CNES, LAM, 38 rue Frédéric Joliot-Curie, 13388 Marseille, France \and
\label{inst:28} SRON Netherlands Institute for Space Research, Niels Bohrweg 4, 2333 CA Leiden, Netherlands \and
\label{inst:29} Gesellschaftsstrasse 6, 3012 Bern, Switzerland \and
\label{inst:31} Leiden Observatory, University of Leiden, PO Box 9513, 2300 RA Leiden, The Netherlands \and
\label{inst:32} Department of Space, Earth and Environment, Chalmers University of Technology, Onsala Space Observatory, 439 92 Onsala, Sweden \and
\label{inst:33} Dipartimento di Fisica, Università degli Studi di Torino, via Pietro Giuria 1, I-10125, Torino, Italy \and
\label{inst:34} National and Kapodistrian University of Athens, Department of Physics, University Campus, Zografos GR-157 84, Athens, Greece \and
\label{inst:35} Astrobiology Research Unit, Université de Liège, Allée du 6 Août 19C, B-4000 Liège, Belgium \and
\label{inst:36} Department of Astrophysics, University of Vienna, Türkenschanzstrasse 17, 1180 Vienna, Austria \and
\label{inst:37} Division Technique INSU, CS20330, 83507 La Seyne sur Mer cedex, France \and
\label{inst:38} Institute for Theoretical Physics and Computational Physics, Graz University of Technology, Petersgasse 16, 8010 Graz, Austria \and
\label{inst:40} ELTE E\"otv\"os Lor\'and University, Institute of Physics, P\'azm\'any P\'eter s\'et\'any 1/A, 1117 Budapest, Hungary \and
\label{inst:41} Lund Observatory, Division of Astrophysics, Department of Physics, Lund University, Box 118, 22100 Lund, Sweden \and
\label{inst:42} IMCCE, UMR8028 CNRS, Observatoire de Paris, PSL Univ., Sorbonne Univ., 77 av. Denfert-Rochereau, 75014 Paris, France \and
\label{inst:43} Institut d'astrophysique de Paris, UMR7095 CNRS, Université Pierre \& Marie Curie, 98bis blvd. Arago, 75014 Paris, France \and
\label{inst:44} Astrophysics Group, Lennard Jones Building, Keele University, Staffordshire, ST5 5BG, United Kingdom \and
\label{inst:45} European Space Agency, ESA - European Space Astronomy Centre, Camino Bajo del Castillo s/n, 28692 Villanueva de la Cañada, Madrid, Spain \and
\label{inst:47} Weltraumforschung und Planetologie, Physikalisches Institut, University of Bern, Gesellschaftsstrasse 6, 3012 Bern, Switzerland \and
\label{inst:48} Dipartimento di Fisica e Astronomia "Galileo Galilei", Università degli Studi di Padova, Vicolo dell'Osservatorio 3, 35122 Padova, Italy \and
\label{inst:49} ETH Zurich, Department of Physics, Wolfgang-Pauli-Strasse 2, CH-8093 Zurich, Switzerland \and
\label{inst:50} Cavendish Laboratory, JJ Thomson Avenue, Cambridge CB3 0HE, UK \and
\label{inst:51} Fachbereich Geowissenschaften, Freie Universität Berlin, Institut für Geologische Wissenschaften, Fachrichtung Planetologie und Fernerkundung, Malteserstr. 74-100, 12249 Berlin \and
\label{inst:52} Institut de Ciencies de l'Espai (ICE, CSIC), Campus UAB, Can Magrans s/n, 08193 Bellaterra, Spain \and
\label{inst:53} Institut d'Estudis Espacials de Catalunya (IEEC), 08860 Castelldefels (Barcelona), Spain \and
\label{inst:54} Institute of Astronomy, University of Cambridge, Madingley Road, Cambridge, CB3 0HA, United Kingdom \and
\label{inst:dlr_main} Deutsches Zentrum für Luft- und Raumfahrt (DLR), Markgrafenstrasse 37, 10117 Berlin, Germany
\and
\label{inst:charn} Universit\'e de Paris, Institut de physique du globe de Paris, CNRS, 75005 Paris, France
\and
\label{inst:beijing} Laboratory for Climate and Ocean-Atmosphere Studies, Department of Atmospheric and Oceanic Sciences, School of Physics, Peking University, Beijing 100871, China
}

   \date{Received \dots; accepted \dots}

 
  \abstract
   {Despite the ever-increasing number of known exoplanets, no uncontested detections have been made of their satellites, known as exomoons.}
   {The quest to find exomoons is at the forefront of exoplanetary sciences. Certain space-born instruments are thought to be suitable for this purpose. We show the progress made with the CHaracterizing ExOPlanets Satellite (CHEOPS) in this field using the HD 95338 planetary system. We present a novel methodology as an important step in the quest to find exomoons.}
   {We utilize ground-based spectroscopic data in combination with Gaia observations to obtain precise stellar parameters. These are then used as input in the analysis of the planetary transits observed by CHEOPS and the Transiting Exoplanet Survey Satellite (TESS). In addition, we search for the signs of satellites primarily in the form of additional transits in the Hill sphere of the eccentric Neptune-sized planet HD 95338b in a sequential approach based on four CHEOPS visits. We also briefly explore the transit timing variations of the planet.}
   {We present refined stellar and planetary parameters, narrowing down the uncertainty on the planet-to-star radius ratio by a factor of $10$. We also pin down the ephemeris of HD 95338b. Using injection/retrieval tests, we show that a $5 \sigma$ detection of an exomoon would be possible at $R_{\rm Moon} = 0.8$~$R_\oplus$ with the methodology presented here.}
   {We exclude the transit of an exomoon in the system with $R_{\rm Moon} \approx 0.6$~$R_\oplus$ at the $1\sigma$ level. The algorithm used for finding the transit-like event can be used as a baseline for other similar targets, observed by CHEOPS or other missions.}


   \maketitle
%

\section{Introduction} \label{sec:intro}

Following the discovery of the first confirmed extrasolar planet orbiting a main sequence star \citep{1995Natur.378..355M}, close to 6000 exoplanets have been identified\footnote{On 11/06/2024, the NASA Exoplanet Archive listed 5787 confirmed planets, \url{https://exoplanetarchive.ipac.caltech.edu/}}. One of the key points of these past decades is the fact that the Solar system is not unique from many aspects, it is simply one of the ever-growing number of known planetary systems in our Galaxy. Based on our understanding of the Solar System, we expect that there are perhaps orders of magnitude more satellites orbiting exoplanets compared to the number of exoplanets. Yet at the time of writing, there are no confirmed exomoons, for which there are a number of possible explanations \citep[see][and references therein]{2024Univ...10..110S}. With the introduction of space telescopes such as Kepler \citep{2010Sci...327..977B} or  CHaracterizing ExOPlanets Satellite \citep[CHEOPS;][]{2021ExA....51..109B} we have the theoretical capability of finding transiting exomoons below the size of Earth. 

The $Kepler$ space telescope has been the first instrument recognized to be able to detect an exomoon \citep{2006A&A...450..395S,2007A&A...470..727S}.
\cite{2015ApJ...813...14K} and \cite{2022NatAs...6..367K} performed extensive analyses of Kepler light curves with the aim of finding transiting exomoons, yielding Kepler-1708b-i as a candidate. Another method thought to be useful for detection of satellites is based on the Transit Timing Variations (TTVs) of the planets \citep{2007A&A...470..727S, 2009MNRAS.392..181K} caused by exomoons. \cite{2021MNRAS.501.2378F} identified 6 candidates based on their Kepler datasets, which were found to be false positives by \cite{2020ApJ...900L..44K}. \cite{2023MNRAS.518.3482K} found no strong TTV signals that could be linked to satellites in a large set of suitable planets observed by Kepler. \cite{2015PASP..127.1084S} predicted that Earth-sized exomoons could likely be detectable with CHEOPS. For that reason, a considerable effort has been devoted within the CHEOPS Science Team to observing those systems which are likely to sustain exomoons for long periods of time \citep[see e.g.][]{2022MNRAS.513.5290D} in pursuit of detecting these objects. \cite{2023A&A...671A.154E} found that there are likely no transiting bodies above the size of Mars orbiting $\nu^2$ Lupi d, based on $1.5$ transits observed with CHEOPS. 

In this paper, we present the necessary steps in the search for moons orbiting the bright ($V = 8.62$~mag) star HD 95338, including refining the stellar and planetary parameters, analysing synthetic data, and placing upper limits on the hypothesized satellite. \cite{2020MNRAS.496.4330D} reported the discovery of a dense Neptune-sized planet ($3.89^{+0.19}_{-0.20}$~$R_\oplus$, $42.44^{+2.22}_{-2.08}$~$M_\oplus$, yielding a density of $3.98^{+0.62}_{-0.64}$~g~cm$^{-3}$), HD 95338b. Based on a single transit observed by Transiting Exoplanet Survey Satellite \citep[TESS;][]{2015JATIS...1a4003R} and a large number of radial velocity measurements, \cite{2020MNRAS.496.4330D} found an orbital period of $55.087 \pm 0.020$ days. This meant that the planet could be far enough from its host star that an exomoon could survive for several $Gyr$ \citep{2022MNRAS.513.5290D}, yet close enough that we can observe it multiple times during the life cycle of CHEOPS, leading to tighter constraints on its parameters, as well as those of its hypothetical companion. However, due to the Sun-synchronous, dusk-dawn low Earth orbit of the satellite, long-period planets might only be observable once a year.

This paper is structured as follows. In Sect. \ref{sec:methods} we describe in detail the photometry of CHEOPS and TESS observations, with a special emphasis on the various detrending steps needed in order to account for the known instrumental effects. We also provide details of the modelling of the transit of HD 95338b. New and improved stellar parameters are provided in Sect. \ref{sec:star}. In Sect. \ref{sec:planet}, we present the refined planetary parameters, based on five transits observed with CHEOPS and TESS. In Sect. \ref{sec:moons}, we present a novel method for chasing transits linked to exomoons. We argue for the need of a thorough noise treatment to avoid false positive exomoon detections. We exclude the presence of Mars-sized transiting exomoons at $1 \sigma$. In Sect. \ref{sec:inj}, we present the analyses of synthetic data, which are used to justify the algorithm used in the search for additional transits. We show that even from a single transit, the $3\sigma$ detection of a $<1$~$R_\oplus$ transit would be feasible in this system.

\section{Methods} \label{sec:methods}
\subsection{CHEOPS photometry} \label{sec:ch_phot}

CHEOPS observed four transits of HD 95338b (during four separate visits) in March 2021, April 2022, May 2023, and April 2025 (Table \ref{tab:visits}). The spacecraft is on a low Earth Orbit, with an $\approx 100$\,min period. Consequently, targets cannot be observed continuously throughout an orbit due to Earth occultations, crossing over the South Atlantic Anomaly and too high stray-light contributions, as is often discussed with regards to CHEOPS data \citep[e.g.][]{2022MNRAS.511.4551L, 2024arXiv240411074K}. This is reflected in the so-called observational efficiency (Table \ref{tab:visits}). We extracted light curves using the CHEOPS Data Reduction Pipeline \citep[DRP version 14.1.2;][]{2020A&A...635A..24H}. The DRP performs the bias, read-out noise, flatfield and smearing-corrections, while also estimating the telescope jitter and background contamination. The pipeline also carries out aperture photometry using a circular apertures, with radii between $15$--$40$ pixels (in 1-pixel steps). In this work, we rely on light curves extracted using the $R = 25$ px radius aperture (the so-called ``default'' mode).

\begin{figure}
    \centering
    \includegraphics[width = \columnwidth]{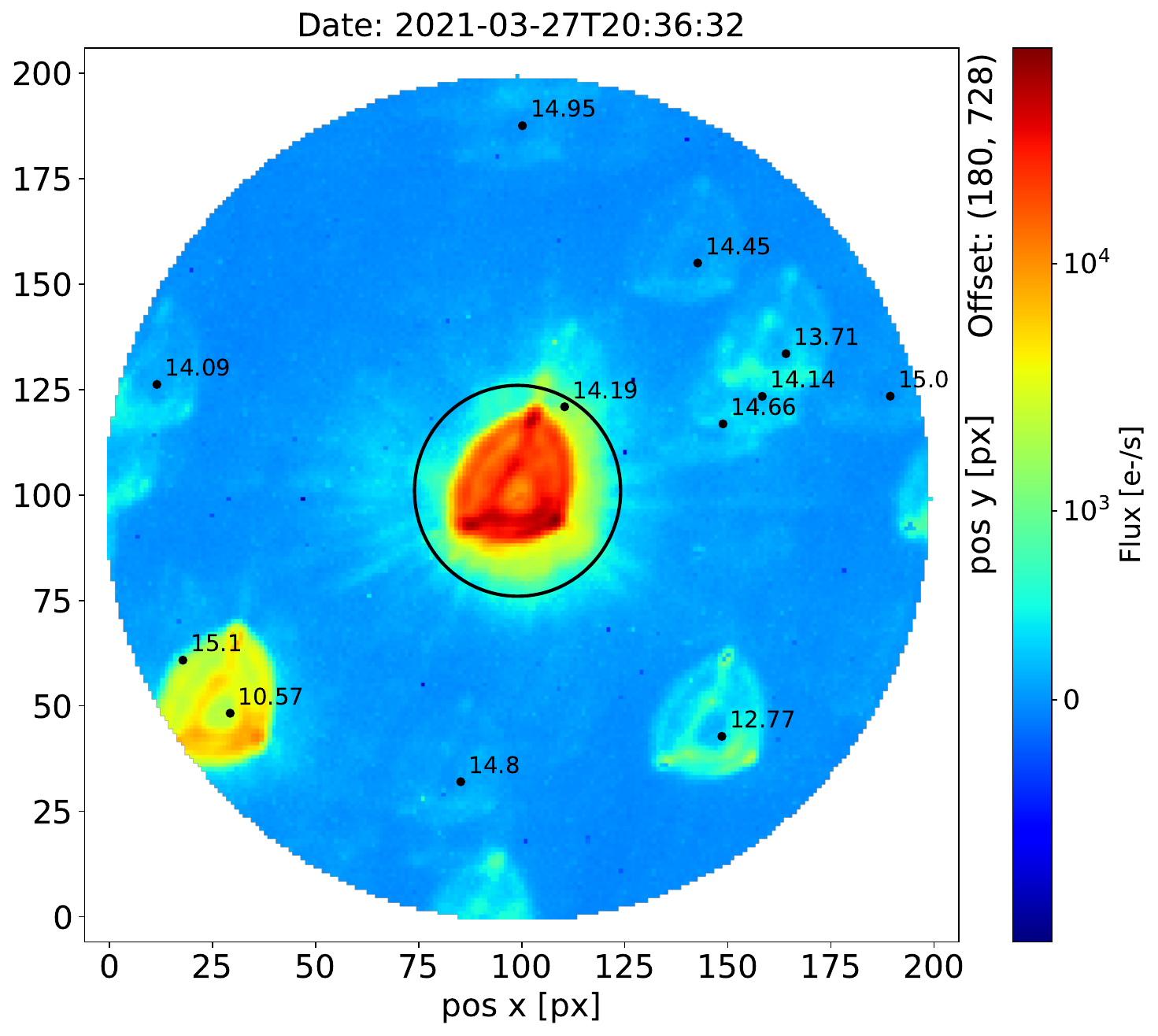}
    \caption{The subarray image depicts the point spread function (PSF) of HD 95338b and the neighbouring stars, with their respective magnitudes of brightness indicated. The target is centred on the coordinates ($180$, $178$) on the CCD.}
    \label{fig:psf}
\end{figure}

Due to the spacecraft rolling around its line of sight, the well-known triangular Point Spread Function \citep[as discussed in e.g.][]{2021A&A...654A.159S, 2023A&A...671A..25K} of any given star (Fig. \ref{fig:psf}) falls onto different pixels throughout each orbit (causing a part of the flux to fall in and out of the circular aperture as the telescope rolls). Several factors, including the inhomogeneous nature of the CCD, stray light, smearing trails and thermal variability of the instrument result in the appearance of periodic or quasi-periodic systematic effects that are correlated with the angular orientation (and possibly the jitter) of the spacecraft. Additionally, the flux contribution of nearby stars also introduces a noise source that changes throughout each orbit and is therefore correlated in nature. The background variation thus also has to be estimated for every exposure. 

We would like to emphasize, that in the context of nominal science observations, a subset of the full frame images, known as "window images", are downlinked. In order to minimise the noise caused by bad pixels, it is necessary to move the position of the window in the CCD to different regions of the CCD when required. In May 2024, the position of the window was relocated to the central location of the CCD\footnote{\url{https://www.cosmos.esa.int/web/cheops-guest-observers-programme/in-orbit-updates}}. This resulted in the fourth visit being observed in a distinct region of the CCD compared to the initial three observations. This modification has resulted in a shift in the configuration of the star pattern within the full frame. This alteration has had an impact on the correlation between the flux and the roll angle of the telescopes (and the number of used harmonics of the roll angle in the analysis). This is due to the fact that the smearing trails of the background stars within the full frame intersect the window as the field rotates around the target star.

\subsection{TESS photometry}

In addition to the CHEOPS photometry described above, we also obtained the short-cadence SAP (Simple Aperture Photometry) light curves from sectors 10 and 63 of TESS \citep[Transiting Exoplanet Survey Satellite;][]{2015JATIS...1a4003R}. These include one transit of HD 95338b each. The transit from sector 10 was also analyzed by \cite{2020MNRAS.496.4330D}.  We corrected for contamination in the aperture using the \verb|CROWDSAP| keyword. We downloaded the light curves from the two sectors from MAST. Additionally, we corrected for the pointing instability of the telescope using the pipeline-estimated position of the photocenter.   As a final preparatory step, we cut out a $1.5$ and a $3$-day window (in sectors 10 and 63, respectively) of the light curves -- that are centered on the two transits -- in order to minimise the computational time needed for the analysis described below.

\subsection{Modelling the transit of HD 95338b from CHEOPS}

There are a number of algorithms and software packages capable of modelling a transiting exoplanet and their satellites, including \verb|planetplanet| \citep{2017ApJ...851...94L}, the phase-folding method by \cite{2021MNRAS.507.4120K},  the analytical framework of \cite{2022ApJ...936....2S}, and \verb|Pandora| \citep{2022A&A...662A..37H}. There are, however, none that incorporate the advanced noise filtering algorithms like those based on Gaussian Processes \citep[GPs; e.g.][]{2006gpml.book.....R, 2012RSPTA.37110550R, 2015arXiv150502965E} or a wavelet-based technique \citep[e.g.][]{2009ApJ...704...51C}. These tools are used to deal with the time-correlated noise arising from both astrophysical and instrumental noise sources. Their benefits in avoiding distortions of the light curve parameters of transiting exoplanets are clearly demonstrated in \cite{2020A&A...634A..75B}, \cite{2023A&A...675A.106C}, and \cite{2023A&A...675A.107K}. Regarding the detection of exomoon transits specifically, \cite{2024MNRAS.528L..66K} and \cite{2024Univ...10..110S} discuss the implications of insufficient noise treatment. 

The two perhaps most well-known exomoon candidates, Kepler-1625b-i \citep{2018SciA....4.1784T} and Kepler-1708b-i \citep{2022NatAs...6..367K}, identified by a so-called photodynamical method, where the parameters of the planet and its satellite are fitted jointly, are not confirmed through independent investigations. In fact, \cite{2019A&A...624A..95H} and \cite{2019ApJ...877L..15K} provide alternative explanation for the exomoon signal observed in Kepler-1625b-i, involving the presence of red noise. More recently, \cite{2024NatAs...8..193H} presented evidence suggesting that ``large exomoons are unlikely'' around both Kepler-1625b and Kepler-1708b, although this is again refuted \citep{2024arXiv240110333K}. It is reasonable to expect that a simultaneous fitting of the planetary and lunar signals with the time-correlated noise would be beneficial in discerning false-positives from bona fide exomoons.

Lacking such a tool, we developed a sequential scheme to search for exomoon-like signals in the CHEOPS light curves of HD 95338b. In the first step, we model the transit of HD 95338b, then in the second step, we search the residuals for signs of additional transits. The transit signal that could be attributed to an exomoon has an amplitude on the order of $\approx 100$\,ppm (as is discussed in Section \ref{sec:moons}), thus a thorough detrending of the known systematic effects is essential to avoid false-positive detections. We therefore construct a strict framework to handle the known (quasi-periodic) systematic effects as follows:

\begin{enumerate}[start = 0]
    \item Remove light curve points that are heavily affected by cosmic rays and remove 4 point before and after each gap (only for the fourth visit)
    \item Preliminary transit modelling of HD 95338b
    \item Find the best-fit model of the instrumental effects in the residuals of the previous step; find the $> 3 \sigma$ outliers in the residuals of this step
    \item Subtract the best-fit systematics model from the raw light curve and mask the outliers identified in the previous step
    \item Model the transit of HD 95338b in the pre-whitened light curve
    \item Check the residuals for transits of an exomoon
\end{enumerate}
In step 2, we construct a basis vector ($F_{\rm BV}$) for detrending \citep[following the common recipe, e.g.][]{2021A&A...654A.159S,2022AJ....164...21S}  in the form of
\begin{equation} \label{eq:systematics}
    F_{{\rm BV}, N} = \sum_{i = 1}^{N} \left( a_i \sin \left( i \phi \right) + b_i \cos \left( i \phi \right) \right),
\end{equation}
where $\phi$ is the roll angle of the spacecraft, $a_i$ and $b_i$ are amplitudes of the sinusoidal terms. We then used the Bayesian Information Criterion \citep[BIC][]{1978AnSta...6..461S} to find the optimal number of harmonics $N$, where $N \in \{1 \dots 10 \}$, to be included in the detrending process. Defining the optimal set of detrending vectors is also part of the optimization problem, and besides the roll angle harmonics, all telemetry data and by-products of the photometry reduction pipeline have initially been considered. We concluded that the background contamination is among the most important detrending parameters, and therefore, we included this vector into the noise model as well.

We analysed the transits of HD 95338b using the Transit and Light Curve Modeller \citep[TLCM;][]{2020MNRAS.496.4442C}. TLCM uses a two-dimensional Gauss-Legendre integration approach to calculate the transit curves. These are described in terms of the scaled semi-major axis of the planet, $a/R_\star$, the planet-to-star radius ratio, $R_{\rm p}/R_\star$, the time of mid-transit, $T_0$, and the orbital period, $P$. \cite{2020MNRAS.496.4330D} found that the planet has an eccentric orbit, specifically, an eccentricity of $e = 0.127 \pm 0.045$ and an argument of the periastron $\omega = 39.4^\circ \pm 18.7^\circ$. In TLCM, it is  possible to account for the orbital eccentricity, where the ingress and egress durations can differ. In that case, instead of the commonly used impact parameter, the so-called (inferior) conjunction parameter contains the information of the orbital inclination of the planet ($i$). It is given by \cite{2020MNRAS.496.4442C} as:
\begin{equation}
    b' = \frac{a}{R_\star} \frac{\left(1-e^2\right) \cos i}{\left( 1+e \sin \omega \right)}.
\end{equation}
We fixed $e$ and $\omega$ at their reported values during the light curve modelling.

Stellar limb darkening is taken into account via the quadratic law \citep{1985A&AS...60..471W}, described by the linear ($u_1$) and quadratic ($u_2$) coefficients using the reparametrization introduced in \cite{2024arXiv240319468K} and \cite{2024arXiv241107797H}:
\begin{align}
    A &= \frac{1}{4}\left( u_1 \left( \frac{1}{\alpha} - \frac{1}{\beta} \right) + u_2 \left( \frac{1}{\alpha} + \frac{1}{\beta} \right) \right), \label{eq:quadldA}\\
    B &= \frac{1}{4}\left( u_1 \left( \frac{1}{\alpha} + \frac{1}{\beta} \right) + u_2 \left( \frac{1}{\alpha} - \frac{1}{\beta} \right) \right) \label{eq:quadldB},
\end{align}
where $\alpha = \frac{1}{2} \cos \left( 77^\circ \right),$ and $\beta = \frac{1}{2} \sin \left( 77^\circ \right)$. In step 1 of the algorithm, the $A$ and $B$ parameters were fitted using the $[1.06,1.56]$ and $[1.40,1.80]$ uniform priors, respectively. These priors are set according to the theoretical limb-darkening of the star (with stellar parameters from Table \ref{tab:star}), as calculated by \texttt{LDCU}\footnote{https://github.com/delinea/LDCU}. This software uses a modified version of the routines of \cite{10.1093/mnras/stv744} to calculated the limb darkening coefficients, based on stellar intensity profiles calculated via the ATLAS \citep{1979ApJS...40....1K} and PHOENIX \citep{2013A&A...553A...6H} models. In step 4, when the definitive planetary parameters are computed, the $A$ and $B$ coefficients were treated as free parameters of the fit.

Furthermore, TLCM uses the wavelet-based formalism of \cite{2009ApJ...704...51C} to handle time-correlated noise (also commonly known as `red noise') that comprises of any remaining instrumental effects or astrophysical variations \citep{2023A&A...675A.106C}. This technique introduces two additional fitting parameters, $\sigma_w$ for the standard deviation of the white noise component and $\sigma_r$ for characterising the red component. These are fitted simultaneously with the parameters describing the transits. To avoid overcorrecting the light curves, the red noise fit is constrained by prescribing that the standard deviation of the white noise (i.e. the residuals) has to be equal to the mean photometric uncertainty of the input light curve \citep{2023A&A...675A.106C}. In step 1, this technique is used to handle the instrumental noise sources, as the wavelet-based noise model is fitted jointly with the transits. Theoretical limb darkening coefficients are needed at this phase to retrieve plausible transit shapes. This allows for a clear detection of the instrument-related effects in step 2. The wavelet-based noise handling is used again in step 4 to obtain the final set of planetary parameters. At this stage, it is expected that astrophysical noise sources will make up the majority of the autocorrelated effects. We emphasize that, from the perspective of the planetary transit fitting, any additional transits -- including those from an exomoon -- are treated as noise by the wavelet technique or other non-parametric methods, and are therefore removed. For this reason, the search for moons needs to occur on a residual light curve where the time-correlated effects (that are not clearly attributed to instrumental effects) are not removed. Therefore, the complex detrending process outlined above is warranted.

\begin{figure*}
    \centering
        \includegraphics[width = \textwidth]{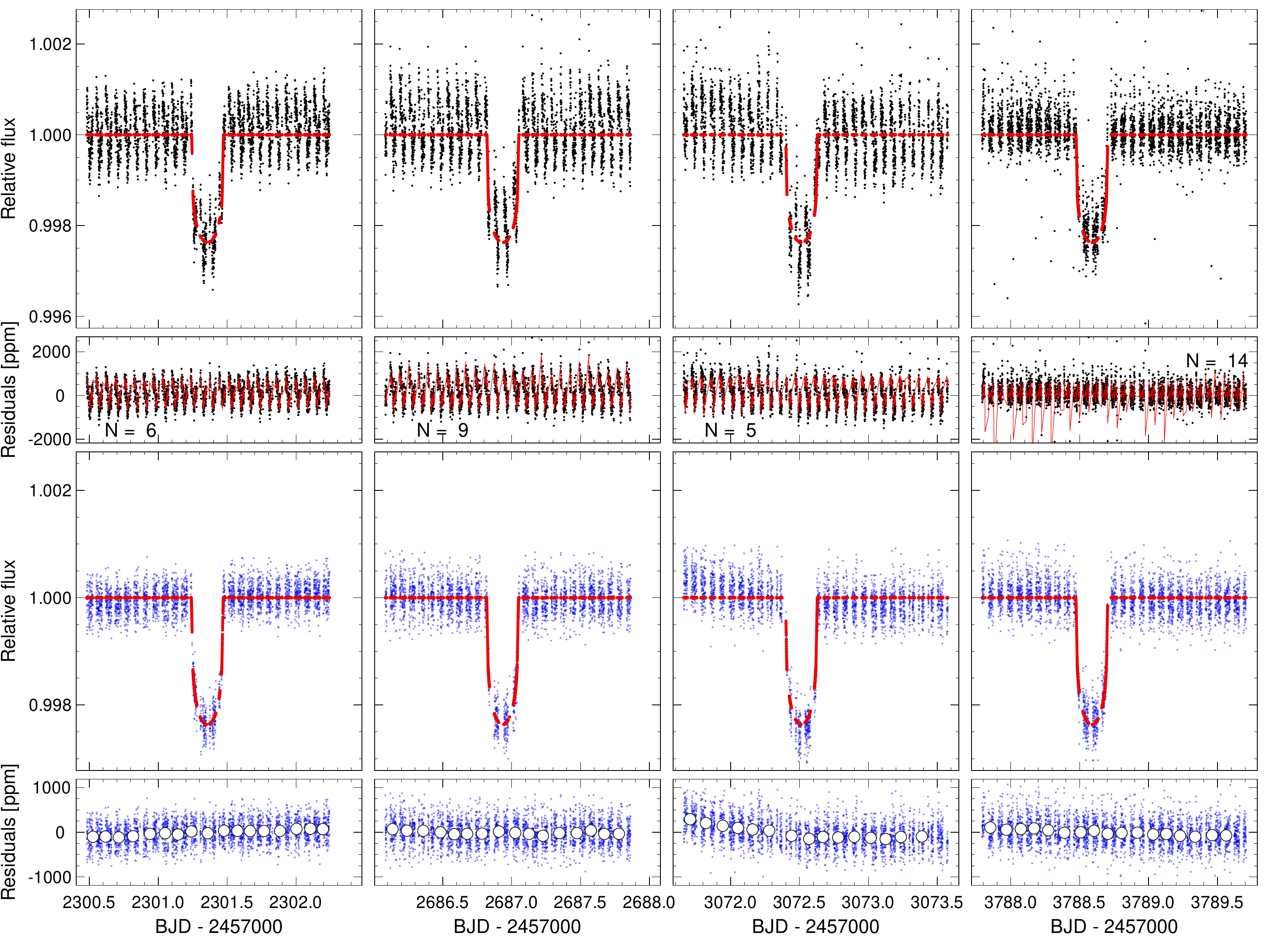}
    \caption{Demonstration of the fitting and detrending steps. The raw CHEOPS light curves of HD 95338b are shown in the top row, along with the initially fitted transit curves (step 1). The residuals of the initial transit fit, along with the best-fit systematics model is shown in the second row, with $N$ number of harmonics of the roll angle included. The final transit fit (step 4) and its residuals (later used in the search for additional transits) are shown in the bottom two rows, where the full white circles represent $200$-point bins (with an average uncertainty of $\approx 20$ ppm in each bin).}
    \label{fig:demo}
\end{figure*}

A much less computationally efficient method would be to fit the $F_{\rm BV}$ jointly with the (planetary) transits and the wavelet-based noise model\footnote{In our experiments, with the available CPUs, the fitting the $F_{\rm BV}$ parameters jointly can be accomplished in $\sim 1$ month, instead of the several hour long runtimes needed with the pre-whitening approach that we implemented.}. This would, however, involve the introduction of up to $66$ additional fitting parameters, since the $a_i$ and $b_i$ coefficients of Eq. (\ref{eq:systematics}), as well as the amplitudes of the background contamination are independent variables in the four visits. In the steps outlined above, the systematic effects are fitted using a simple least squares method, which is several orders of magnitude faster than the joint $F_{\rm BV}$ plus red noise plus transit fit in our experience. This way, we are able to drastically narrow down the parameter space of the transit modelling itself, thus reducing the possibilities for the MCMC chains to get stuck in so-called local minima, and increasing the reliability of the solutions.

\subsection{Joint TESS+CHEOPS modelling}

In order to increase the precision of the recovered planetary parameters, we also performed a joint TESS+CHEOPS LC analysis. We assume that the passband-independent orbital parameters, including the eccentricity, the argument of periastron, the semi-major axis and the orbital period remain constant on the timescale relevant in these observations. We also assume that although the TESS passband is redder compared to CHEOPS \citep[see e.g.][]{2024arXiv240916268S}, the possible atmospheric features on HD 95338b are not detectable with the present noise levels, meaning that there is no wavelength-dependence of the transit depth. Therefore, we assume that $R_p/R_\star$ is the same in the two passbands. We further assume that the transit chord does not change between the different observations, meaning that the scaled semi-major axis, the conjunction parameter, and the orbital period are the same for all five modelled transits. The limb-darkening coefficients are passband-dependent, which neccesitates the inclusion of $A^{\rm CHEOPS}$, $B^{\rm CHEOPS}$, $A^{\rm TESS}$ and $B^{\rm TESS}$ into the analysis. It is understood that red noise from an astrophysical origin decreases with increasing wavelength. This manifests in an expectedly lower $\sigma_r$ \citep[see e.g.][]{2023MNRAS.522..488K} in the TESS observations compared to the CHEOPS light curves. On the other hand, the white noise level depends primarily on the aperture size and the exposure time. We therefore include $\sigma_r^{\rm CHEOPS}$, $\sigma_w^{\rm CHEOPS}$, $\sigma_r^{\rm TESS}$, and $\sigma_w^{\rm TESS}$ in the global LC modelling. The likelihood of the joint TESS+CHEOPS modelling is then the sum of the log-likelihoods of the light curve fits of the data from the two satellites.

In addition to increasing the number of transits that can be analyzed, the two different passbands also allow us to break the degeneracy between the limb-darkening coefficients and the transit parameters, yielding more precise estimates on both.

During the revision of this paper, \cite{2025MNRAS.539..928S} published an independent analysis based on the first three transits from CHEOPS and the two TESS transits. 

\subsection{Sampling the posterior distributions}

In every TLCM-based light curve analysis described in this paper, we followed the same recipe. We used wide, uninformed priors on every parameter. A genetic algorithm is then able to find the (global) minimum in the likelihood space, with $N_{\rm init} = 2000$ randomly selected values for each parameter \citep[see ][for more details]{2020MNRAS.496.4442C}. Having found starting points that are suspected to be good, we applied a Differential Evolution Markov-chain Monte Carlo \citep[DE-MCMC;][]{2006S&C....16..239T} to get robust estimates of the uncertainty ranges of the fitting parameters. In every case, we used $10$ chains (also known as walkers) each consisting of $20000$ possible steps. The convergence was monitored via the Gelman-Rubin statistic \citep{1992StaSc...7..457G} and the effective sample size \citep[e.g.][]{Ripley87}. If convergence is not reached within these $20000$ steps, the chain length is automatically extended to $10 \cdot 20000$.

\section{Stellar parameters} \label{sec:star}

\begin{table}[]
    \caption{Stellar parameters of HD 95338.}
    \label{tab:star}
    \centering
    \begin{tabular}{l c c}
    \hline
    \hline
        Parameter & This work & \cite{2020MNRAS.496.4330D}\\
        \hline
        $T_{\rm eff}$ [K] & $5097 \pm 77$ & $5212^{+16}_{-11}$\\
        $\log g$ [cgs] & $4.44 \pm 0.04$ & $4.54 \pm 0.21$\\
        $[Fe/H]$ & $0.08 \pm 0.05$ & $0.04 \pm 0.10$ \\
        microturbulence [km s$^{-1}$] & $0.60 \pm 0.12$ & -- \\
        $v_\star \sin i_\star$ [km s$^{-1}$] & $1.23 \pm 0.23$ & $1.23 \pm 0.28$\\
        $R_\star$ [$R_\odot$] & $0.868 \pm 0.006$ & $0.87 \pm 0.04$\\
        $M_\star$ [$M_\odot$] & $0.848 \pm 0.047$ & $0.83 \pm 0.02$ \\
        $t_\star$ [Gyr] & $9.5^{+4.3}_{-5.7}$ & $5.08 \pm 2.51$ \\
        \hline
    \end{tabular}
\end{table}

We used the ARES+MOOG spectroscopic analysis methodology to derive the stellar spectroscopic parameters ($T_{\mathrm{eff}}$, $\log g$, microturbulence, [Fe/H]). This methodology is described in detail in \citet[][]{Sousa-21, Sousa-14, Santos-13}. The equivalent widths (EW) were consistently measured using the ARES code\footnote{The last version, ARES v2, can be downloaded at https://github.com/sousasag/ARES} \citep{Sousa-07, Sousa-15}. We used the list of iron lines appropriate for stars cooler that 5200 K which was presented in \citet[][]{Tsantaki-2013}. For this spectral analysis we used a combined HARPS spectrum of HD 95338. To converge for the best set of spectroscopic parameters for each spectrum we use a minimization process to find the ionization and excitation equilibrium. This process makes use of a grid of Kurucz model atmospheres \citep{Kurucz-93} and the latest version of the radiative transfer code MOOG \citep{Sneden-73}. We also derived a more accurate trigonometric surface gravity using recent GAIA data following the same procedure as described in \citet[][]{Sousa-21} which provided a consistent value (4.44 $\pm$ 0.04 dex) when compared with the spectroscopic surface gravity.

We computed the stellar radius of HD 95338 utilising a MCMC modified infrared flux method \citep[IRFM --][]{Blackwell1977,Schanche2020} that allows us to produce synthetic photometry from a constructed spectral energy distribution (SED) using stellar atmospheric models \citep{Castelli2003} and our stellar spectroscopic parameters as priors. These simulated data are compared to broadband fluxes in the following bandpasses: \textit{Gaia} $G$, $G_\mathrm{BP}$, and $G_\mathrm{RP}$, 2MASS $J$, $H$, and $K$, and WISE $W1$ and $W2$ \citep{Skrutskie2006,Wright2010,GaiaCollaboration2022} that are listed in Table 4 of \cite{2020MNRAS.496.4330D} to derive the stellar bolometric flux. The last step in the MCMC is to convert this flux to the stellar effective temperature and angular diameter, that is combined with the offset-corrected \textit{Gaia} parallax \citep{Lindegren2021} to produce the stellar radius. 

To accurately characterise the stellar properties of HD95338 and account for some modelling systematics, we derived two pairs of $\left\{M_\star, t_\star\right\}$ from two different approaches. The first mass and age pair were estimated by applying the isochrone placement algorithm \citep{Bonfanti2015_stellar_age, Bonfanti2016_stellar_age} that interpolates the input stellar parameters within pre-computed grids of isochrones and tracks of \texttt{PARSEC}\footnote{\textsl{PA}dova and T\textsl{R}ieste \textsl{S}tellar \textsl{E}volutionary \textsl{C}ode: \url{http://stev.oapd.inaf.it/cgi-bin/cmd}} v1.2S \citep{Marigo2017_PARSEC}. 
The second pair of mass and age estimates were derived from a Levenberg-Marquardt minimisation scheme \citep[as detailed in ][]{Salmon2021_AlphaCen} computing stellar models on the fly with the \texttt{CL\'ES} code \citep[Code Li\`egeois d'\'Evolution Stellaire --][]{Scuflaire2008_CLES}. This minimisation scheme optimises over the age and mass to produce the best agreement between model and observed radii and effective temperatures. As detailed in \citet{Bonfanti2021_HD108236}, we checked the mutual consistency between the two respective pairs of outcomes via a $\chi^2$-based criterion and then merged the results obtaining $M_\star=0.848 \pm 0.043~M_\odot$ and $t_\star=9.5^{+4.3}_{-5.7}$~Gyr. 

The derived stellar parameters are listed in Table \ref{tab:star}. In general, they are in good agreements with the values reported by \cite{2020MNRAS.496.4330D}, although we present more precise estimates on $R_\star$ and $\log g$, which are used in the transit modelling.

In order to increase the precision and accuracy of the planetary model, we incorporated the stellar parameters of Table \ref{tab:star} as inputs to the light curve analysis. By a rearrangement of Kepler's third law of planetary motion, the scaled semi-major axis is connected to the stellar density as well, since (under the assumption of a spherical star)
\begin{equation}
    \left( \frac{a}{R_\star} \right)^3 = P^2 G \left(1+q \right) \frac{\rho_{\star \rm, transit}}{3 \pi},  
\end{equation}
where $q = M_{\rm p}/M_\star$, and $\rho_{\star, \rm transit}$ is a geometrically constrained density, since $a/R_\star$ is determined from the transit duration. Using the estimated $T_{\rm eff}$, [Fe/H] and $\rho_{\star, \rm ransit}$, we are able to provide estimate $M_{\star, \rm transit}$ and $R_{\star, \rm transit}$ using the empirical formulae\footnote{We assume that [Fe/H] = [M/H].} of \cite{2011MNRAS.417.2166S}. Then, according to \cite{2020MNRAS.496.4442C}, the goodnes-of-fit metric (in this case, the logarithmic likelihood) can be modified as

\begin{equation} \label{eq:likelihood_star}
\begin{split}
    -\log \mathcal{L}_{\rm mod} = -\log \mathcal{L} &+ \frac{1}{2} \left( \frac{R_\star-R_{\star, \rm transit}}{\sqrt{\left(\Delta R_\star\right)^2+\left( \Delta R_{\star, \rm transit} \right)^2}} \right)^2 \\ &+ \frac{1}{2} \left( \frac{\log g- \log g\vert_{\rm transit}}{\sqrt{\left(\Delta  \log g\right)^2+\left( \Delta \log g\vert_{\rm transit} \right)^2}} \right)^2,
\end{split}    
\end{equation}
where $\log g\vert_{\rm transit} = 4.43 - 2 \log R_{\star, \rm transit} + \log M_{\star, 
 \rm transit}$, where $\log g_\odot = 4.43$. Given that both the second and third terms contain information about $\rho_{\rm transit}$, $a/R_\star$ is constrained via the input stellar parameters as well.

\section{Results} \label{sec:planet}
\subsection{Planetary parameters} \label{sec:planetparams}

\begin{figure}
    \centering
    \includegraphics[width = \columnwidth]{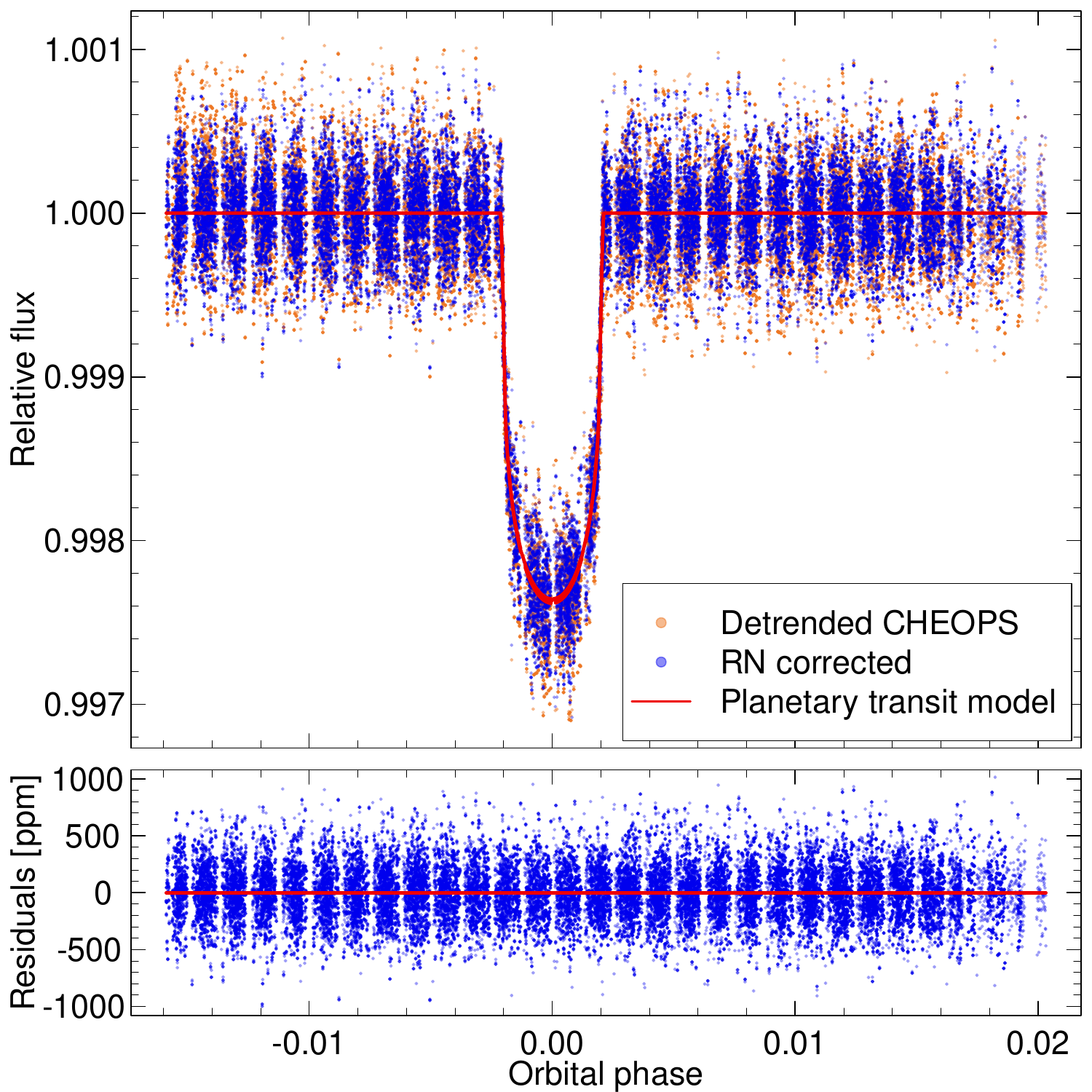}
    \caption{Phase-folded detrended CHEOPS light curve of HD 95338 (top panel, orange, the same data as the third row of Fig. \ref{fig:demo}.) The rednoise-corrected light curve is shown with blue, while the best-fit transit light curves of HD 95338b is represented by the solid red curve.}
    \label{fig:transit_ch}
\end{figure}

\begin{figure}
    \centering
    \includegraphics[width = \columnwidth]{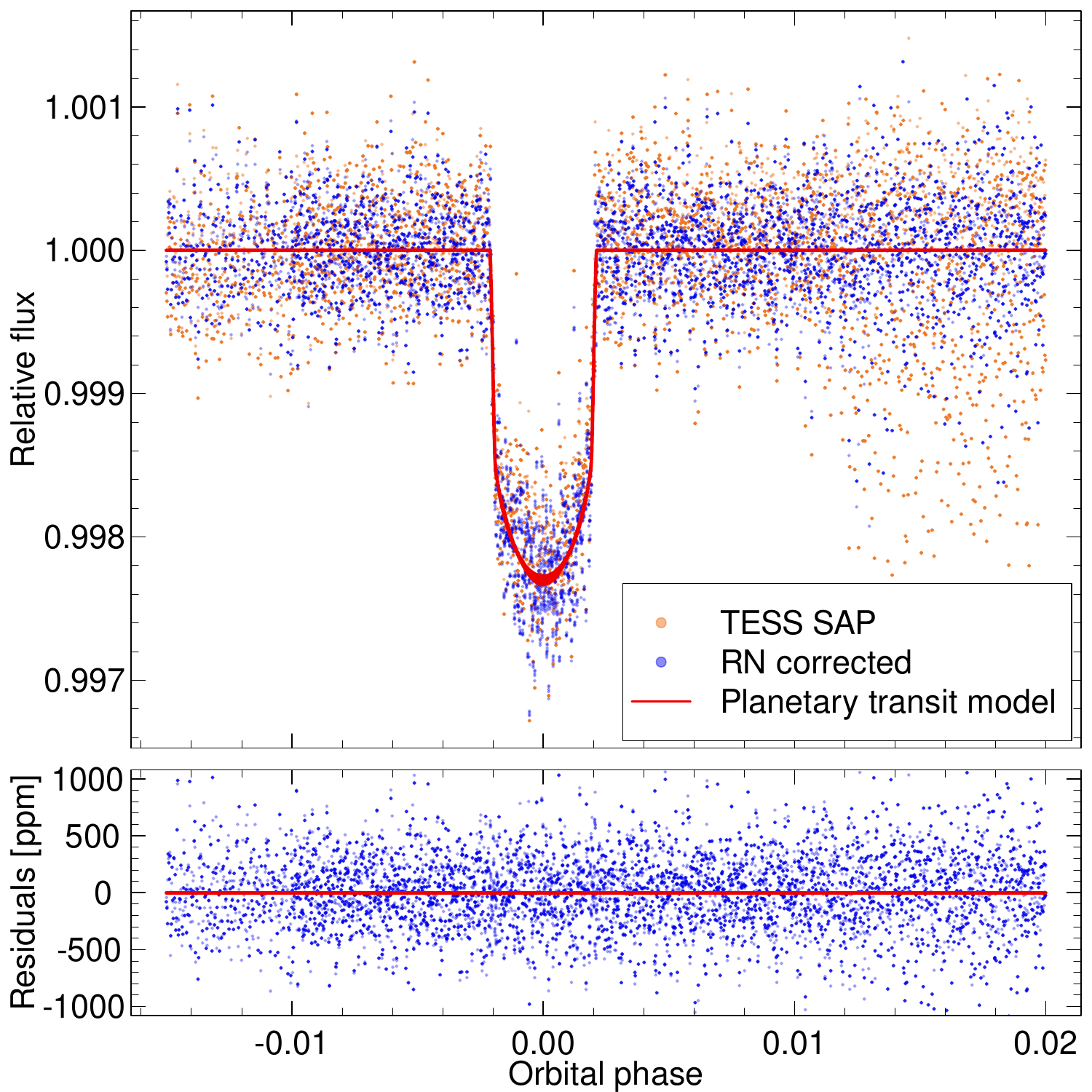}
    \caption{Phase-folded detrended TESS light curve of HD 95338 (top panel, orange, the same data as the third row of Fig. \ref{fig:demo}.) The rednoise-corrected light curve is shown with blue, while the best-fit transit light curves of HD 95338b is represented by the solid red curve.}
    \label{fig:transit_t}
\end{figure}

The best-fit CHEOPS and TESS light curves (from the joint fit) are shown in Figs. \ref{fig:transit_ch} and \ref{fig:transit_t}. The corresponding parameters are shown in Table \ref{tab:params_comp_sep_sec}. The parameters describing the planetary transit agree well with each other between the CHEOPS only and the joint TESS plus CHEOPS modelling. We note that the uncertainty ranges do not decrease with the inclusion of the two additional TESS transits, which implies that they are dominated by the uncertainty in the input stellar parameters. The best-fit conjunction parameter is in a $<1 \sigma$ agreement with the impact parameter presented by \citep{2020MNRAS.496.4330D}, and the relative planetary radius shows a discrepancy of only $\approx 1.2 \sigma$. There is a $4.8 \sigma$ disagreement between the estimated $a/R_\star$ values, and a $2.8 \sigma$ difference between the $T_0$ from our joint TESS+CHEOPS analysis and the respective values shown by \citep{2020MNRAS.496.4330D}. Using the S/N criteria established by \cite{2023A&A...675A.106C} for $a/R_\star$, we know that from the CHEOPS analysis alone, we are able to constrain the scaled semi-major axis well within 2\% of the truth (since for a 2\% accuracy, a S/N = $3$ is needed, and in our case, S/N = $22$). Consequently, we argue that the uncertainty range presented in \citep{2020MNRAS.496.4330D} is likely underestimated. \cite{2020MNRAS.496.4330D} find that the first TESS transits occurs $166 \pm 52$ seconds earlier than our estimates, which is likely a combination of inadequate noise treatment \citep{2023A&A...671A..25K} and the fact that we analyse five transits together. However, the discrepancy in $T_0$ is $< 3\sigma$. A brief TTV (transit timing variation) analysis is shown in Sect. \ref{sec:ttv}. 

The smaller aperture of the TESS cameras implies that the TESS observations are more noisy -- this is confirmed by the higher $\sigma_w^{\rm TESS}$ compared to $\sigma_w^{\rm CHEOPS}$. The red noise parameters are not directly comparable between the two instruments, although we expect the this noise type to decrease at longer wavelengths \citep[e.g. ][]{2023MNRAS.522..488K}. The retrieved $A^{\rm CHEOPS}$ and $B^{\rm CHEOPS}$ are within $3 \sigma$ of the theoretical values. Both limb darkening coefficients in the TESS passband are in agreement with those presented by \cite{2020MNRAS.496.4330D} (after conversion to the formalism used in this paper).

Our results are also in agreement with the parameters presented by \cite{2025MNRAS.539..928S}. Due to the additional transit included in this work and the refined stellar parameters, we estimate slightly lower uncertainties in $R_p/R_\star$, $b'$, $P$, and $T_0$. A direct comparison of the confidence intervals of $a/R_\star$ and the limb-darkening coefficients is not applicable, since \cite{2025MNRAS.539..928S} did not use these as fitting parameters.

We find that the planetary orbit has an inclination of $89.654^\circ \pm 0.0155^\circ$, the planet has a radius of  $R_p = 0.3811 \pm 0.0035 R_{\rm Jup}$, with a transit duration of  $5.587 ^{+0.018} _{-0.016}$ hours.

\begin{table*}
\scriptsize
\caption{Parameters obtained from the transit modelling of HD 95338b.}
\label{tab:params_comp_sep_sec}
\centering
\begin{tabular}{l c c c c c}
\hline
\hline
 Parameter & Prior & CHEOPS & CHEOPS + TESS (global)&  \cite{2020MNRAS.496.4330D} & \cite{2025MNRAS.539..928S}\\
\hline
$a/R_\star$ & [$1.0$, $133.0$] & $67.19 \pm 0.40$ &  $67.34 \pm 0.40$ 
 & $64.676^{+0.381}_{-0.384}$ & $69.1_{-2.3}^{+1.8}$\\
$R_p/R_\star$ & [$0$, $0.2$] & $0.04423 \pm 0.00027$ &  $0.04413  \pm 0.00026$
 & $0.0409 \pm 0.0028$ & $0.04369^{+0.00049}_{-0.00041}$\\
$b'$ & [$0$, $1.0$] & $ 0.373 \pm 0.018$ & $0.370  \pm 0.016$ 
& $0.430^{+0.070}_{-0.113}$ & $0.25^{+0.11}_{-0.15}$\\
$T_0$ [BJD - 245700] & [$2300.35$, $2302.35$] &  $2301.35590 \pm 0.00033$ &  $1585.28087  \pm   0.00050$ \tablefootmark{a}
&$1585.2795 \pm 0.0006$ & $1585.28072^{+0.00055}_{-0.00054}$\\
$P$ [days] & [$54.056$, $56.056$] & $55.082697 \pm 0.000020$ &  $55.082694 \pm 0.000018$
& $55.087 \pm 0.020$ & $55.082695 \pm 0.000027$\\
$\sigma_r^{\rm CHEOPS}$ & [$0$, $0.1$] &   $0.00552 \pm 0.00019$
 &  $0.00553 \pm   0.00019$  &  -- & --\\
$\sigma_w^{\rm CHEOPS}$ &[$0$, $0.1$] &  $0.0002265 \pm 0.0000015$ 
 & $0.000226  \pm  0.0000015$ &-- & --\\
 $\sigma_r^{\rm TESS}$ & [$0$, $0.1$] &   -- 
 &  $0.0068 \pm   0.00028$ &   -- & --\\
$\sigma_w^{\rm TESS}$ &[$0$, $0.1$] &  -- 
 & $0.0003727  \pm  0.0000050$ &-- & --\\
$A^{\rm CHEOPS}$ & [$-1.5$, $2.5$] & $1.51 \pm  0.24$ & $1.47 \pm  0.22$ & -- & $1.2 \pm 2.1$\\
$B^{\rm CHEOPS}$ & [$-2.0$, $2.0$]& $1.747 \pm  0.062$ & $1.747 \pm  0.061$ &-- & $1.1 \pm 2.0$\\
$A^{\rm TESS}$ & [$-1.5$, $2.5$] & -- & $0.53 \pm  0.18$  & $0.73\pm 0.59$ & $1.2 \pm 2.3$\\
$B^{\rm TESS}$ & [$-2.0$, $2.0$]& -- & $1.44 \pm  0.15$ & $1.467 \pm 0.25$ & $1.1 \pm 2.3$\\
\hline
\end{tabular}
\tablefoot{\tablefoottext{a}{Uniform prior applied: $[1585.2790, 1585.2800]$.}}
\end{table*}

\subsection{Transit timing} \label{sec:ttv}

\begin{figure}
    \centering
    \includegraphics[width=\columnwidth]{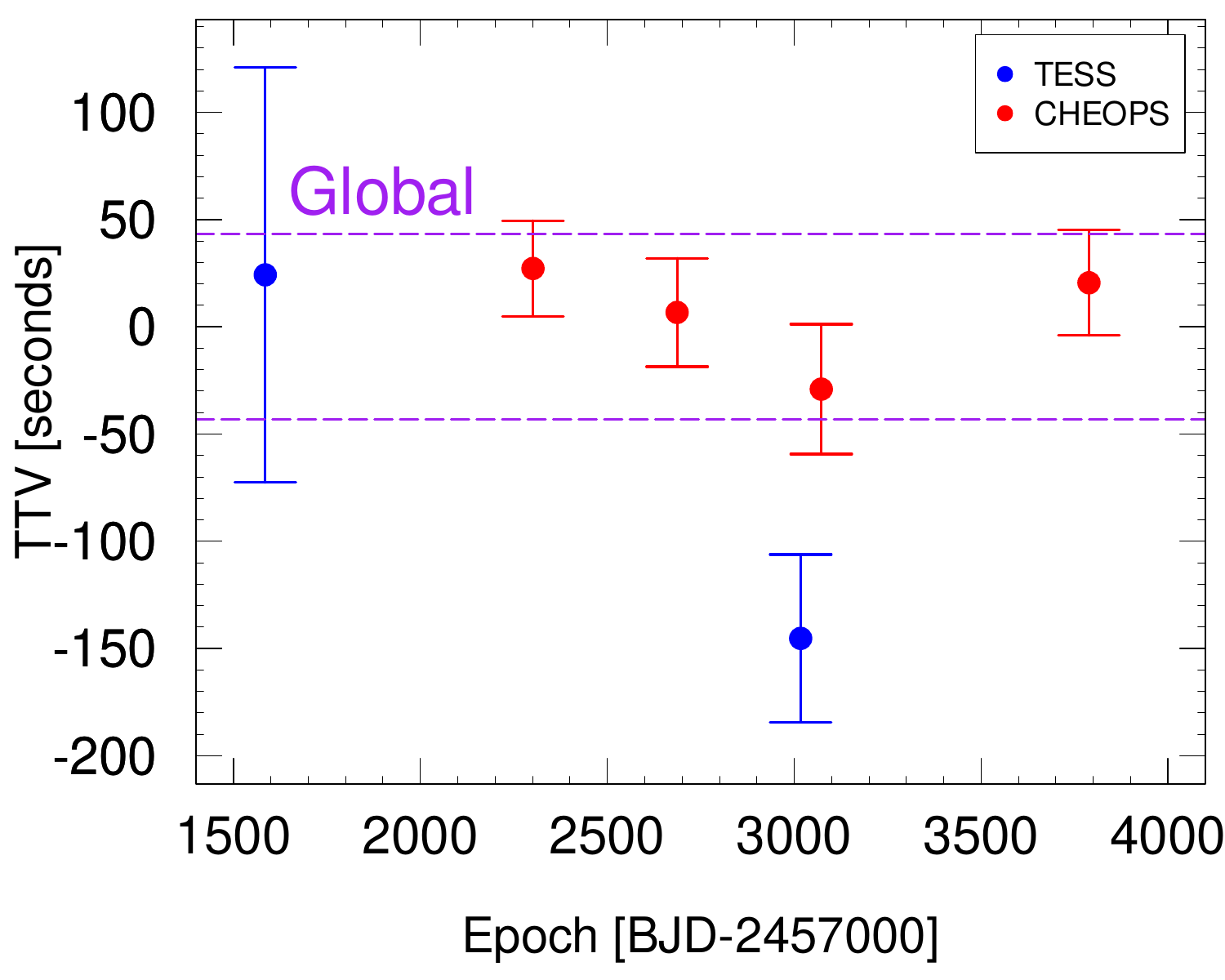}
    \caption{Transit timing data of HD 95338b. The dashed purple lines represent the uncertainty range from the global fit.}
    \label{fig:ttv}
\end{figure}

We extracted the timing of each individual transit (two from TESS,  four from CHEOPS) by fixing all parameters to the best-fit values from Table \ref{tab:params_comp_sep_sec} with the exception of $\sigma_w$, $\sigma_r$ and $T_0$, assuming a constant orbital period. We also included a possible quadratic trend in the modellings. The resultant transit timing data are shown in Fig. \ref{fig:ttv} and are compiled in Table \ref{tab:ttv}. There is no apparent TTV, meaning that the transit timings are explained well by a constant period. The $T_0$ value from sector 63 is in a $2.5 \sigma$ disagreement with the global modelling, which is not statistically significant. A deeper exploration of the TTV signals is beyond the scope of this paper.

\begin{table*}
\scriptsize
\caption{Transit timing data (and other selected parameters from the transit modelling) of HD 95338b.}
\label{tab:ttv}
\centering
\begin{tabular}{l c c c c c c}
\hline
\hline
Instrument & $t_c$ & $p_0$ [ppm] & $p_1$ [ppm] & $p_2$ [ppm] & $\sigma_r$ [ppm] & $\sigma_w$ [ppm]  \\
\hline
CHEOPS & $2301.35620 \pm 0.00026$ & $-4.5 \pm 11.7$ & $117.9 \pm 12.2$ & $-13.3 \pm 24.7$ & $611.5 \pm 259.5$ & $211.8 \pm 2.5$ \\ 
CHEOPS & $2686.93482 \pm 0.00029$ & $-32.0 \pm 22.5$ & $-51.2 \pm 20.2$ & $83.6 \pm 41.0$ & $1252.2 \pm 257.4$ & $236.4 \pm 2.9$ \\ 
CHEOPS & $3072.51326 \pm 0.00035$ & $-44.9 \pm 38.1$ & $-278.4 \pm 29.5$ & $172.7 \pm 57.0$ & $2218.3 \pm 290.7$ & $230.5 \pm 3.3$ \\ 
CHEOPS & $3788.58886 \pm 0.00028$ & $-4.4 \pm 13.9$ & $-112.5 \pm 15.3$ & $43.8 \pm 25.8$ & $742.8 \pm 435.3$ & $259.1 \pm 3.2$ \\ 
TESS & $1585.28114 \pm 0.00112$ & $-301.9 \pm 98.2$ & $-742.0 \pm 141.4$ & $517.8 \pm 114.9$ & $4439.6 \pm 342.6$ & $418.0 \pm 8.0$ \\ 
TESS & $3017.42922 \pm 0.00045$ & $57.2 \pm 15.4$ & $231.1 \pm 10.1$ & $-36.2 \pm 12.0$ & $690.5 \pm 397.9$ & $257.4 \pm 3.0$ \\ 
\hline
\end{tabular}
\end{table*}

\section{The search for moons} \label{sec:moons}

We chose to restrict the search for exomoons on the CHEOPS dataset alone. This is done because the smaller (mirror) aperture of TESS in comparison to CHEOPS manifests as higher white noise levels, which are detrimental to the quest for moons with the methodology presented here ($191$~ppm for CHEOPS compared to $289$~ppm in TESS despite the exposure time that is more than 4 times shorter, Table \ref{tab:params_comp_sep_sec}). For that reason, and to keep the analysis self-consistent, we use the planetary orbital parameters ($P$, $b$ and $a/R_\star$) that are derived from the CHEOPS photometry alone in the processes described below.

\subsection{Sensitivity maps}
\begin{figure*}
    \centering
    \includegraphics[width = \textwidth]{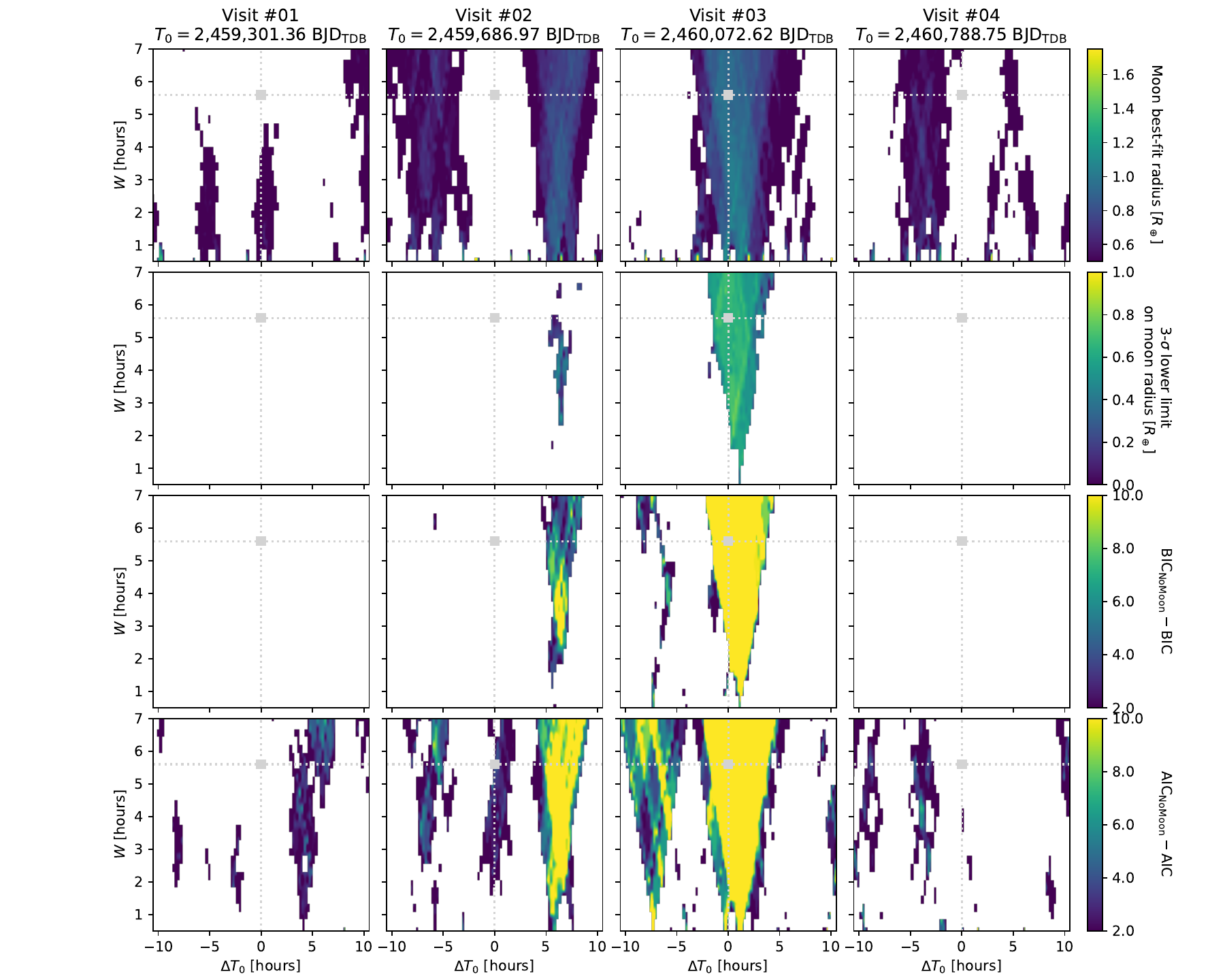}
    \caption{Sensitivity maps for the four CHEOPS visits of HD 95338. The first row represents the best-fit box depth (i.e. moon radius) in the $\Delta T_0$--$W$ space. The $3\sigma$ lower limit of the moon radius is shown in the second row, while the third and fourth rows represent the $\Delta$BIC and $\Delta$AIC values between a solution with and without a moon.
    The empty (blank) areas are values that are below a threshold of $0\,R_\oplus$ for the moon radius (first 2 rows) and a threshold of 2 for the $\Delta$BIC and $\Delta$AIC (last 2 rows).
    The grey squares (and corresponding dotted lines) show the position of the transit of HD 95338b in this parameter space for each visit. An underlying linear trend is assumed for each of the four visits.}
    \label{fig:sensitivity-maps}
\end{figure*}
We examined the CHEOPS residual light curves for shallow transit signals (that could be linked to an exomoon) following the approach detailed in \cite{2023A&A...671A.154E}. This method explores a grid of moon transit epochs and durations $\left\{T_{0, \textrm{Moon}},W\right\}$. For each point of the grid, we fit for a box-shaped transit depth (allowed to be negative) together with a trend (linear or quadratic) and a multiplicative noise term (factor applied to the data error bars). The fit is done using the Markov chain Monte Carlo algorithm (MCMC) implemented in the Python package \texttt{emcee} \citep{2013PASP..125..306F}. From the posterior distribution on the transit depth, we derive the best-fit value of the moon radius and its 3-$\sigma$ lower limit. We also compute the Bayesian Information Criterion and the Akaike Information Criteria \citep[AIC;][]{Akaike1998} defined as:
\begin{align}
    \textrm{BIC} &= k \ln\!\left(n\right) - 2\ln\!\left(\mathcal{L}\right) \label{eq:bic}\\
    \textrm{AIC} &= 2k - 2\ln\!\left(\mathcal{L}\right),
\end{align}
where $n$ is the number of data points, $k$ is the number of parameters included in the fit, $\mathcal{L}$ is the best-fit (maximum) likelihood value. The BIC and AIC values are compared to their respective values in the case where the moon transit depth is fixed to 0, i.e. a `no-moon' reference fit.

The parameter space scanned by the $\left\{T_{0, \textrm{Moon}},W\right\}$ grid must cover mid-transit times across the full Hill-sphere of \mbox{HD 95338b}\footnote{The Hill sphere defines the gravitational sphere of influence of a body and is therefore where moons are expected to exist. \cite{2006MNRAS.373.1227D} shows that beyond half the Hill sphere radius, only retrograde orbits are stable, and we cover the full radius not to exclude these potential objects.} while allowing for transit duration either shorter or longer (e.g. if the moon impact parameter is larger or smaller) than that of its planet. With a transit duration of the Hill sphere of $17.6\pm0.9$\,hours (i.e. $<20.3$\,h at 3\,$\sigma$), we considered $T_{0, \textrm{Moon}}$ values that satisfy $-10.5\,\textrm{h} \leq \Delta T_0 \leq +10.5\,\textrm{h}$, where $\Delta T_0 = T_{0, \textrm{Moon}}-T_{0, \textrm{Planet}}$. The range of moon transit durations $W$ covers values from 30\,min (grazing moon) to 7\,h, with a planetary transit duration of $\sim$\,5.6\,h. The grid resolution (increment) is set to 15\,min in both dimensions.

In order to provide conservative estimates of a possible exomoon signal in the $T_0$ -- $W$ parameter space, we propagate the uncertainties from the planetary transit fit. To that end, we calculate the standard deviation of $100$ red noise plus planetary transit (see Sect. \ref{sec:planetparams}) models (with parameters selected randomly from the posterior distribution of the fit) at every time stamp of the observations, and add this to the flux uncertainties estimated by DRP in quadrature. In order to increase computational efficiency, we fit $F_{\rm BV}$ via a least-squares algorithm. Consequently, we do not propagate the uncertainties of $F_{\rm BV}$ into the residuals.

\subsection{Box-shift method} \label{sec:boxshift}

We performed a more detailed transit modelling after the planetary transit in the third visit to try to explore the possibility of an additional transit being there. To do that, we used narrow box priors on $T_{0, {\rm Moon}}$ which allowed us to perform an in-depth scan of the selected areas, by shifting them through the light curve. The priors are chosen to be $3$ hours wide ($\approx 1/2$ transit duration), and at each new step, their borders are increased by $1.5$ hours ($\approx 1/4$ transit duration), yielding an overlapping configuration for the transit modelling. In total, we performed $8$ different transit modellings in every light curve.

We assume that the inclination of the orbit of the moon is such that the impact parameter is the same as for HD~95338b. We further assume that the separation of the planet and its companion is so small that the star--moon distance can be described sufficiently well with the semi-major axis of the planetary orbit, and that the relative velocity of the moon is negligible compared to the orbital velocity of the system. The difference in the transit duration, although difficult to estimate as it depends on the mass of the moon and its semi-major axis in the orbit around the planet. Based on the specific configuration, it is possible that a moon within the Hill sphere would not be transiting or would show a grazing transit \citep{2023A&A...671A.154E}, or that the duration of the lunar transit would exceed that of the planet \citep{2024MNRAS.528L..66K}. It is also established that the limb-darkening coefficients can more reliably be measured in case of deeper transits \citep[e.g. Eq (34) of][]{2023A&A...675A.106C}. For these reasons, we only fit the depth and epoch of the lunar transit, while we adopt all other transit parameters from the analysis of the planetary transits (Table \ref{tab:params_comp_sep_sec}).

\subsection{Injection-and-retrieval tests} \label{sec:inj}

In order to verify the performance of the sequential exomoon-detecting algorithm presented above, we performed a number of injection-and-retrieval tests. We injected transits with the  $a/R_\star$ and $b$ values of \citet{2020MNRAS.496.4330D}. We tested a wide range of moon radii, with a grid consisting of \mbox{$R_{\rm Moon, injected} \in \{ 0.8, 0.9, 1.05, 1.2, 1.35\}R_\oplus$}. The transits were injected at two epochs (BJD $2459301.25$ in the first visit and BJD $2459687.05$ in the second visit) so that they overlap with the transit of HD 95338b. We followed the same steps as described in Sect. \ref{sec:ch_phot}, then proceeded to generate the sensitivity maps, and, having identified the transits on these we also performed the box-shifting approach as well. There is an infinite number of possible $T_{0, \textrm{Moon}} - R_{\rm Moon}$ combinations that could be used in such tests. These specific cases are useful to demonstrate that the technique described above works well for finding lunar transits. 
We only use data from the first three CHEOPS visits for the injection-and-retrieval tests.

In Figs. \ref{fig:drp25_def16_senmap_quad} and \ref{fig:drp25_def27_senmap_quad} we show the sensitivity maps for the $R_{\rm Moon, injected} = 0.8$~$R_\oplus$ and $R_{\rm Moon, injected} = 1.35$~$R_\oplus$ cases. The second rows, showing the $3\sigma$ lower limit on the moon radii show on clear strip for about $4$ hours before and $5$ hours after the conjunction of HD 95338b in the first and second visits, respectively. The presence of the signal is also confirmed by the $\Delta$BIC and $\Delta$AIC values (third and fourth rows of Figs. \ref{fig:drp25_def16_senmap_quad} and \ref{fig:drp25_def27_senmap_quad}, respectively). The $\Delta$BIC and $\Delta$AIC metrics also show that at certain times ($\approx 2$ hours after and $\approx 1$ hour before the midtransit of the planet) positive bumps (i.e. negative moon radii) are preferred to the combination of GPs and the quadratic trend. Such signals do not correspond to the transit of hypothetical exomoon. Comparing with Fig. \ref{fig:drp25_def0_senmap_quad}, we may conclude that they appear as a consequence of the injected lunar transits. It is likely that these two additional signals act as weights on the planetary transit, causing minor distortions that are then detected during the sensitivity mapping.

After identifying these regions of interest (between $\approx 2301.04$ and $\approx 2301.26$ for the first visit, $\approx 2686.93$ and $\approx 2697.15$ for the second visit), we also conducted the box-shifting parameter extraction similarly to the study of the real dataset (Sec. \ref{sec:boxshift}). We include a quadratic trend (Eq. \ref{eq:quad}) in the analyses. For the cases of $R_{\rm Moon, injected} = 0.8$~$R_\oplus$ and $R_{\rm Moon, injected} = 1.35$~$R_\oplus$, the light curves from this method are shown on Figs. \ref{fig:boxshift_injections_def16} and \ref{fig:boxshift_injections_def27}. We can observe the transit appearing and then disappearing again as the searchbox for $T_{0, \textrm{Moon}}$ is shifted through the light curves.

The extracted $T_{0, \textrm{Moon}}$ and $R_{\rm Moon}$ values are listed in Tables \ref{tab:recovery_visit1} and \ref{tab:recovery_visit2} for all tested moon sizes, along with the respective deviations from the nominal values. The transit depths are recovered even for $R_{\rm Moon, injected} = 0.8$~$R_\oplus$ case in both visits. We find that the significance of this parameter (and consequently the transit itself) is $\approx 5.5$~$R_\oplus$ in the first visit and $\approx 4.2$ in the second visit for $R_{\rm Moon, injected} = 0.8$~$R_\oplus$. The significances of the detected transits naturally increase with $R_{\rm Moon, injected}$. Table \ref{tab:ttv} suggests that in the second visit, the time-correlated noise level is higher at $\sigma_r = 1356.0 \pm 259.8$~ppm compared to just  $\sigma_r = 589.8 \pm 255.2$~ppm. We emphasize that anything not explicitly included in the modelling (via a fitting parameter) is described by the wavelets. For this reason, $\sigma_r$ depends heavily on the underlying light curve (whether it contains injected transits or not) and the actual fitted quantities. Consequently, comparison of the red noise levels between the various $R_{\rm Moon, injected}$ cases is not possible.
The $\approx 900$~ppm difference in the red noise levels may explain the larger uncertainties on $R_{\rm Moon}$ seen in the second visit. Similarly, the in the third visit, we find $\sigma_r = 2443.8 \pm 282.3$~ppm (Table \ref{tab:ttv}), which may shroud any evidence for an exomoon. A comprehensive analysis of the underlying causes of these discrepancies is beyond the scope of this paper, but it should be noted that the instrument displays a certain degree of sensitivity loss in relation to its ageing process \citep[][Sect. 4]{2024A&A...687A.302F}.

We note that the new window on the CCD yielded considerably lower red noise (Table \ref{tab:ttv}).

Although one of the injected moon transits is on the ingress of the planetary transit and the other one is on the egress, they are not symmetric with respect to $T_0$. Consequently, the shape of the modelled planetary transit gets distorted in comparison with the original dataset, and this discrepancy is more evident the larger the injected exomoon radius is. In this sequential approach of the chase for moons, we are in fact fitting the photocenter \citep{2007A&A...470..727S} as the ``planet''. The retrieved ``planetary'' parameters are therefore more distorted if we increase $R_{\rm Moon, injected}$. This is shown in Table \ref{tab:planetparams_injections}. We observe a shift in $T_0$ to earlier epochs, and increase in $R_{\rm p}/R_\star$, and a decrease in $a/R_\star$.  Although the recovered parameters agree in a $1 \sigma$ with each other and even the original dataset (Table \ref{tab:boxshift-params}), these minor inconsistencies may lead to differences in the residuals that are then converted into the parameter biases seen in Tables \ref{tab:recovery_visit1} and \ref{tab:recovery_visit2}. Furthermore, the additional signals (i.e. transits) can also cause and increased variance in the $100$ models chosen from the posteriors for the uncertainty propagation. As a result, the hint of an exomoon seen on the left column Fig. \ref{fig:drp25_def0_senmap_quad} disappears (left columns of Fig. \ref{fig:drp25_def16_senmap_quad}, \ref{fig:drp25_def27_senmap_quad}). A more detailed investigation of the distortion of the planetary parameters is beyond the scope of this paper.

Given that the distortion of the transit parameters and the flux uncertainties increases with $R_{\rm Moon, injected}$, the recovered $R_{\rm Moon}$ and $T_{0, \rm Moon}$ are also further from their input values at the higher moon radii (Table \ref{tab:recovery_visit1}, \ref{tab:recovery_visit2}). At the same time, the lower moon sizes yield $\{T_{0, \rm Moon}, R_{\rm Moon}\}$ combinations that are within $1 \sigma$ of the injections.

\begin{table*}
\caption{Retrieved planetary parameters for the different injected exomoon sizes.}
\label{tab:planetparams_injections}
\centering
\begin{tabular}{l c c c c}
\hline
\hline
Size of injected moon [$R_\oplus$]& $T_0$ & $R_{
\rm p}/R_\star$ & $a/R_\star$ & b' \\
\hline
$R_{\rm Moon} = 0.80$ & $2301.35602 \pm 0.00031$ & $0.04435 \pm 0.00026$ & $67.74 \pm 0.40$ & $0.344 \pm 0.019$ \\ 
$R_{\rm Moon} = 0.90$ & $2301.35600 \pm 0.00032$ & $0.04437 \pm 0.00026$ & $67.76 \pm 0.41$ & $0.345 \pm 0.019$ \\ 
$R_{\rm Moon} = 1.05$ & $2301.35594 \pm 0.00032$ & $0.04444 \pm 0.00027$ & $67.73 \pm 0.41$ & $0.344 \pm 0.019$ \\ 
$R_{\rm Moon} = 1.20$ & $2301.35589 \pm 0.00032$ & $0.04453 \pm 0.00027$ & $67.73 \pm 0.40$ & $0.345 \pm 0.019$ \\ 
$R_{\rm Moon} = 1.35$ & $2301.35585 \pm 0.00032$ & $0.04462 \pm 0.00027$ & $67.71 \pm 0.40$ & $0.345 \pm 0.018$ \\ 
\hline
\end{tabular}
\end{table*}

\section{The search for moons orbiting HD 95338b}
\begin{figure*}
    \centering
    \includegraphics[width=\textwidth]{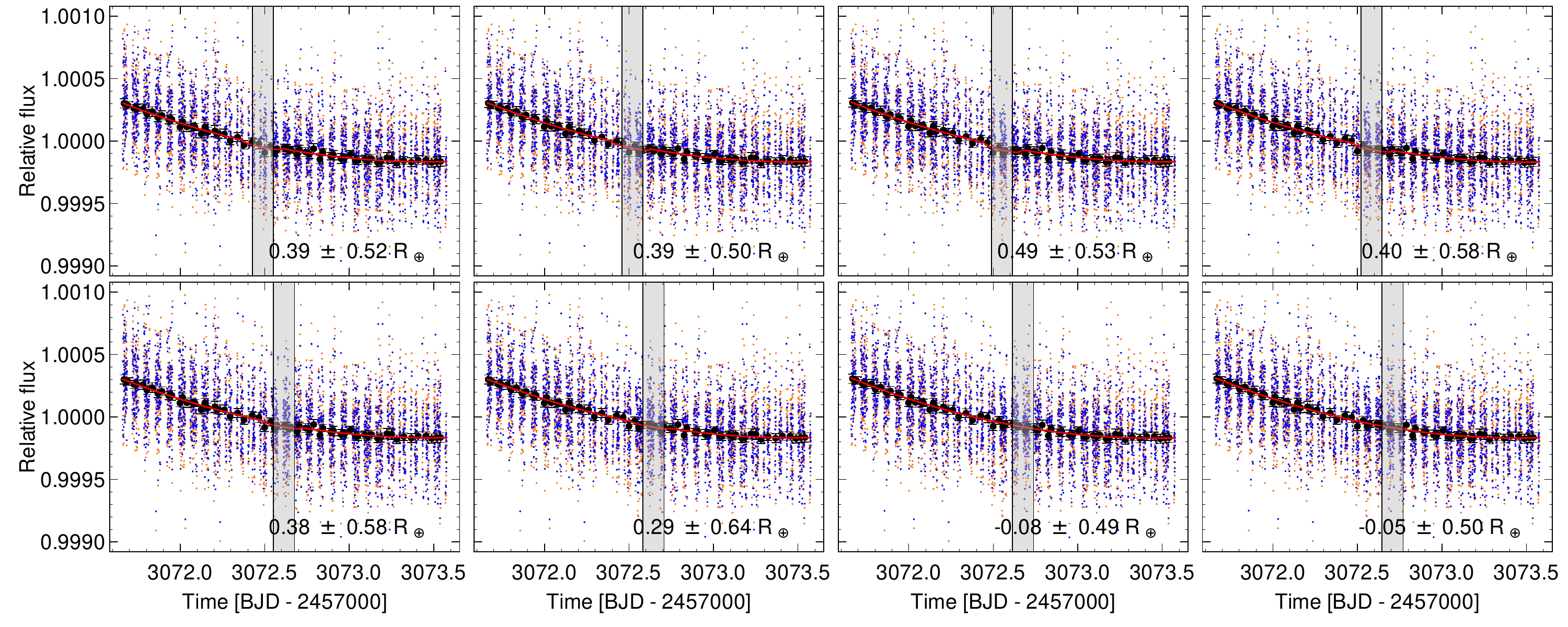}
    \caption{Results of the box-shifting method. The shaded gray areas represent the searchboxes within which the mid-time of the lunar transit is constrained. The best-fit moon radius and its $1 \sigma$ uncertainty is shown on the bottom of each plot.}
    \label{fig:boxshift_quad}
\end{figure*}
Having verified that we are able to achieve $>3 \sigma$ detections of additional transiting bodies with $0.8$~$R_\oplus$ based on single transits (Sect. \ref{sec:inj}), we explored the residual light curves of HD 95338 from the four CHEOPS visits. The sensitivity map for each visit of \mbox{HD 95338} is shown in Fig. \ref{fig:sensitivity-maps} (with an underlying linear trend), which hints and the presence of a signal that occurs slightly later than the planetary transit during the second visit (middle column), and a strong signal in the third visit (right column). The signals in these two visits can be described by a box with a depth corresponding to $0.8$~$R_\oplus$ and $\approx 1.2$~$R_\oplus$, respectively. We also estimate 3-$\sigma$ lower limits at  $\approx 0.6$~R$_\oplus$ (in the second visit) and $\approx 0.8$~R$_\oplus$ (in the third visit). We find that the cases where a box-shaped signal is included in the fit is favoured by $\Delta \text{BIC} \geq 10$ and $\Delta \text{AIC} \geq 10$. In the first visit (Fig. \ref{fig:sensitivity-maps}, left column), the lower limit on the hypothetical transit depth is negative at every ($\Delta T_0$, $W$) grid point. In the case of the second transit, the detection statistics are more sporadically distributed in the parameter space and not in a manner that we expect for a transit based on the injection--and--retrieval tests (see Sect. \ref{sec:inj}).

The spread-out detections (non-zero $3\sigma$ lower limits) before the transit in the second visit are likely noise. In the first and fourth visits, the best-fit box does not have a $3 \sigma$ detection significance (i.e. the $3\sigma$ lower limit on the radius is negative), and the inclusion of the box is disfavoured by the BIC. For these reasons, we conclude that a moon signal is not detectable in the first two visits or the last visit, while the signal seen in the third visit warrants further investigations. We note that at the bottom edge of the maps (Fig. \ref{fig:sensitivity-maps}) the width of the boxes (i.e. the duration of the supposed lunar transit) is commensurable with the duration of the gaps in observations, as discussed in Sect. \ref{sec:ch_phot}. For this reason, any signal seen there can also safely be disregarded.

We also constructed the sensitivity maps by using a quadratic trend (Fig. \ref{fig:drp25_def0_senmap_quad}). Firstly, we note that no signals appear in the first or fourth visits. Secondly, we observe that the features seen in the second and third visits (Fig. \ref{fig:sensitivity-maps}) disappear, as the $3\sigma$ lower limit (second row of Fig. \ref{fig:drp25_def0_senmap_quad}) no longer implies a clear detection. The BIC and AIC values still suggest that the inclusion of a dip is preferred, although lower with probabilities ($\Delta$BIC $\approx 8$ instead of $\geq 10$ as seen in the case when a linear trend is included).

\begin{table*}
\caption{Temporal coordinates of the eight searchboxes used in the box-shifting analysis of the exomoon signal seen in the third visit, and the best-fit parameters describing the correlated noise, the transit and the long-term trend.}
\label{tab:boxshift-params}
\centering
\scriptsize
\begin{tabular}{l c c c c c c c c c}
\hline
\hline
Searchbox number & Searchbox left & Searchbox right & $T_{0, \textrm{Moon}}$ & $R_{\rm Moon}/R_\star$ & $\sigma_r$ [100 ppm] & $\sigma_w$ [100 ppm] & $p_0$ [ppm] & $p_1$ [ppm]  & $p_2$ [ppm]\\
\hline
1 &  $3072.4265$ & $3072.5515$ & $3072.482 \pm 0.040$ & $0.0041 \pm 0.0055$ & $30.9 \pm 2.0$ & $2.234 \pm 0.030$ & $-27 \pm 54$ & $-284 \pm 42$ & $145 \pm 75$ \\ 
2 &  $3072.4578$ & $3072.5828$ & $3072.513 \pm 0.045$ & $0.0040 \pm 0.0052$ & $30.9 \pm 2.0$ & $2.234 \pm 0.030$ & $-27 \pm 54$ & $-284 \pm 42$ & $145 \pm 75$ \\ 
3 &  $3072.4890$ & $3072.6140$ & $3072.563 \pm 0.043$ & $0.0052 \pm 0.0056$ & $30.9 \pm 2.0$ & $2.233 \pm 0.030$ & $-27 \pm 54$ & $-284 \pm 42$ & $145 \pm 75$ \\ 
4 &  $3072.5203$ & $3072.6453$ & $3072.591 \pm 0.035$ & $0.0042 \pm 0.0061$ & $30.9 \pm 2.0$ & $2.235 \pm 0.030$ & $-27 \pm 54$ & $-284 \pm 42$ & $145 \pm 75$ \\ 
5 &  $3072.5515$ & $3072.6765$ & $3072.599 \pm 0.036$ & $0.0040 \pm 0.0061$ & $30.9 \pm 2.0$ & $2.235 \pm 0.030$ & $-27 \pm 54$ & $-284 \pm 42$ & $145 \pm 75$ \\ 
6 &  $3072.5828$ & $3072.7078$ & $3072.623 \pm 0.044$ & $0.0031 \pm 0.0067$ & $30.9 \pm 1.9$ & $2.234 \pm 0.030$ & $-27 \pm 54$ & $-284 \pm 42$ & $145 \pm 75$ \\ 
7 &  $3072.6140$ & $3072.7390$ & $3072.676 \pm 0.041$ & $-0.0008 \pm 0.0052$ & $30.9 \pm 2.0$ & $2.234 \pm 0.030$ & $-27 \pm 54$ & $-284 \pm 42$ & $145 \pm 75$ \\ 
8 &  $3072.6453$ & $3072.7703$ & $3072.705 \pm 0.043$ & $-0.0006 \pm 0.0052$ & $31.0 \pm 2.0$ & $2.235 \pm 0.030$ & $-27 \pm 54$ & $-284 \pm 42$ & $145 \pm 75$ \\ 
\hline
\end{tabular}
\end{table*}

After narrowing down the possible locations of the exomoon transit via the sensitivity maps (Fig. \ref{fig:sensitivity-maps}), we pinned down the exact position and duration of the signal. We identified the relevant light curve section to be between $\approx 246072.427$ BJD and $\approx 246072.777$ BJD. We constructed eight overlapping segments (each spanning $3$ hours, or roughly half of the transit duration of HD 95338b). These segments are used as uniform priors (or searchboxes) for mid-time of the lunar transit. To scan the above-mentioned portion of the light curve, we shift these searchboxes by $45$ minutes (or one-eight of the planetary transit duration) for each individual fit. We thus perform eight individual fits in search of the lunar transit, as seen on Fig. \ref{fig:boxshift_quad}. We also included a quadratic trend (for the flux--time detrending) in the form of
\begin{equation}
    F_{\rm trend} = p_0 + p_1 \cdot \left( t - T_{\rm ref} \right) + p_2 \cdot \left(t-T_{\rm ref} \right)^2,
\end{equation} \label{eq:quad}
where $p_0$, $p_1$, and $p_2$ are the constant, linear, and quadratic coefficients respectively, and $t_{\rm ref} = T_{0, \rm Moon}$ is the reference time of the trend. We also allow for the two parameters of the wavelet-based noise filtering to vary in an unconstrained manner. The results of the box-shifting method are shown on Fig. \ref{fig:boxshift_quad}, and the best-fit parameters are listed in Table \ref{tab:boxshift-params}. The noise-related parameters ($\sigma_r$, $\sigma_w$, $p_0$, $p_1$ and $p_2$) are consistent with each other in all eight cases, while $R_{\rm Moon}/R_\star$ is consistent with $0$ within $1 \sigma$. We therefore do not see a significant detection of a transit-like dip.

\section{Discussion and concluding remarks}

In this work, we carried out the search for possible satellites orbiting the eccentric Neptune-sized planet HD 95338b in a sequential approach. Using four transits from CHEOPS and two from TESS, we present a detailed transit modelling, followed by a novel approach in the quest for exomoons. We re-derived the stellar parameters of the host star HD 95338, and present considerable improvements on some of them in comparison with the discovery paper \citep{2020MNRAS.496.4330D}. We present improvements on the basic transit parameters of HD 95338b. Most notably, we reduce the uncertainty on the planetary radius by a factor of $10$ compared to the results presented by \cite{2020MNRAS.496.4330D}. Given that the precision of these parameters does not change with the inclusion/exclusion of the two TESS transits, we argue that the stellar parameters and their uncertainties are properly taken into account in TLCM \citep[as described in][]{2020MNRAS.496.4442C} and that the uncertainties of the modelling parameters are limited by our knowledge of the star. The most remarkable improvement is achieved regarding the orbital period, which we estimate with a precision of $\approx 0.3$ s compared to $\approx 1700$ s in the discovery paper.

The quest for exomoons is difficult as we are trying to find transits whose timing and duration are not known, and that are shallow (especially in comparison with the instrumental noise effects of CHEOPS). A thorough detrending, incorporating all known noise sources is therefore a pre-requisite for such a project, as outlined in Sect. \ref{sec:methods}. We scanned the residual light curves after subtracting the planetary signal for signs of additional transit-like features that could be attributed to an exomoon orbiting HD 95338b. For a thorough characterization of an exomoon which would incorporate the combination of two Keplerian orbits \citep{2022A&A...662A..37H}, at least three transit detections would be needed. We are able to rule out the presence of additional signals below $\approx 0.6$~$R_\oplus$ at $1 \sigma$, thus placing an upper limit on the size of a hypothetical satellite in the system.

There are obvious long-term trends in the residuals seen in Fig. \ref{fig:demo}. With the assumption that these are linear (and letting the complex noise handling tools, either GP-based or wavelet-based, handle all other possible non-white noise realizations), we see the detection of a transit-like feature (Fig. \ref{fig:sensitivity-maps}). However, with the addition of a quadratic trend (Figs. \ref{fig:drp25_def0_senmap_quad} and \ref{fig:boxshift_quad}), the transit detection is no longer significant. The deviation between the fitted linear and quadratic trends is the greatest around the position of the detected transit. The fact that the significance of the detection decreases after including a higher-order time-dependent term -- which is commonly associated with astrophysical noise sources, such as spots -- implies that the transit-like feature is likely not originated by a transit of an exomoon. Increasing the order of polynomials included in the modelling introduces the risk that the shorter time-scale noise components are enhanced in a way that may introduce additional transit-like features, thus we do not test for these possibilities, relying instead on the noise handling algorithms that are thought to be more efficient at higher frequencies. In any case, the retrieved moon size is consistent with a no-moon case at $3 \sigma$. Additionally, the lack of obvious TTVs (Fig. \ref{fig:ttv}) also serves as counterargument against the presence of a satellite \citep{2007A&A...470..727S, 2009MNRAS.392..181K}.

Based on the analysis of the synthetic LCs, we conclude that the method presented in Sect. \ref{sec:moons} is viable for searching for exomoon transits in a non-photodynamical way. The injection-and-retrieval tests suggest that the combination of the sensitivity maps and the box-shifting method does not induce false positive detections, however, it does recover the size and position of the injected exomoon transits (within $2\sigma$ of their respective input values), and leads to $>3 \sigma$ detections even for the smallest tested moon sizes. 

HD 95338b is an interesting target and an important milestone in the quest for exomoons. Future observations with CHEOPS, PLATO \citep[PLAnetary Transits and Oscillations of stars][]{2014ExA....38..249R, 2024arXiv240605447R}, Ariel \citep[Atmospheric Remote-sensing Infrared Exoplanet Large-survey][]{2018ExA....46..135T, 2022EPSC...16.1114T}, or JWST \citep[James Webb Space Telescope][]{2006SSRv..123..485G} may further reduce the current upper limit of $0.6$~$R_\oplus$ on the radius of a possible moon in the system. The methodology presented in Sections \ref{sec:methods} and \ref{sec:moons} can also be regarded a stepping stone in the hunt for exomoons using the telescopes of the future. In the event of detections of additional signals that are similar to the one seen on Fig. \ref{fig:boxshift_quad}, a fully photodynamic characterisation may be able to provide deeper insights into the nature of this system.

The discovery of the first confirmed exomoon will open a new chapter in the history of exoplanet research. The above told tale about the exomoon hunt around HD 95338b, together with the earlier examples mentioned in Section \ref{sec:intro}, sends a strong message to the research community regarding this subject. Detecting exomoons is currently situated at the frontiers of contemporary research, and success in confirming such discoveries can only be achieved through a comprehensive understanding and in-depth exploration of the intricacies involved in signal design, measurement techniques, and data analysis. The discussion also undertakes to outline the potential pathways to be explored during this investigation, and we are convinced that our considerations in the above analysis will provide significant methodological aspects to this demanding task.

 
\begin{acknowledgements}
CHEOPS is an ESA mission in partnership with Switzerland with important contributions to the payload and the ground segment from Austria, Belgium, France, Germany, Hungary, Italy, Portugal, Spain, Sweden, and the United Kingdom. The CHEOPS Consortium would like to gratefully acknowledge the support received by all the agencies, offices, universities, and industries involved. Their flexibility and willingness to explore new approaches were essential to the success of this mission. CHEOPS data analysed in this article will be made available in the CHEOPS mission archive (\url{https://cheops.unige.ch/archive_browser/}). This work has made use of data from the European Space Agency (ESA) mission {\it Gaia} (\url{https://www.cosmos.esa.int/gaia}), processed by the {\it Gaia} Data Processing and Analysis Consortium (DPAC,
\url{https://www.cosmos.esa.int/web/gaia/dpac/consortium}). Funding for the DPAC
has been provided by national institutions, in particular the institutions
participating in the {\it Gaia} Multilateral Agreement. On behalf of the ``Searching for transiting exomoons in CHEOPS data'' project SzK is grateful for the possibility to use HUN-REN Cloud \citep[see][https://science-cloud.hu/]{H_der_2022} which helped us achieve the results published in this paper.

ASi and CBr acknowledge support from the Swiss Space Office through the ESA PRODEX program. 
A.De. 
GyMSz acknowledges the support of the Hungarian National Research, Development and Innovation Office (NKFIH) grant K-125015, a a PRODEX Experiment Agreement No. 4000137122, the Lend\''ulet LP2018-7/2021 grant of the Hungarian Academy of Science and the support of the city of Szombathely. 
This project has received funding from the Swiss National Science Foundation for project 200021\_200726. It has also been carried out within the framework of the National Centre of Competence in Research PlanetS supported by the Swiss National Science Foundation under grant 51NF40\_205606. The authors acknowledge the financial support of the SNSF. 
TWi acknowledges support from the UKSA and the University of Warwick. 
S.G.S. acknowledge support from FCT through FCT contract nr. CEECIND/00826/2018 and POPH/FSE (EC). 
The Portuguese team thanks the Portuguese Space Agency for the provision of financial support in the framework of the PRODEX Programme of the European Space Agency (ESA) under contract number 4000142255. 
The Belgian participation to CHEOPS has been supported by the Belgian Federal Science Policy Office (BELSPO) in the framework of the PRODEX Program, and by the University of Liège through an ARC grant for Concerted Research Actions financed by the Wallonia-Brussels Federation. 
GSc, LBo, VNa, IPa, GPi, RRa, and TZi acknowledge support from CHEOPS ASI-INAF agreement n. 2019-29-HH.0. 
ABr was supported by the SNSA. 
YAl acknowledges support from the Swiss National Science Foundation (SNSF) under grant 200020\_192038. 
We acknowledge financial support from the Agencia Estatal de Investigación of the Ministerio de Ciencia e Innovación MCIN/AEI/10.13039/501100011033 and the ERDF “A way of making Europe” through projects PID2019-107061GB-C61, PID2019-107061GB-C66, PID2021-125627OB-C31, and PID2021-125627OB-C32, from the Centre of Excellence “Severo Ochoa” award to the Instituto de Astrofísica de Canarias (CEX2019-000920-S), from the Centre of Excellence “María de Maeztu” award to the Institut de Ciències de l’Espai (CEX2020-001058-M), and from the Generalitat de Catalunya/CERCA programme. 
DBa, EPa, and IRi acknowledge financial support from the Agencia Estatal de Investigación of the Ministerio de Ciencia e Innovación MCIN/AEI/10.13039/501100011033 and the ERDF “A way of making Europe” through projects PID2019-107061GB-C61, PID2019-107061GB-C66, PID2021-125627OB-C31, and PID2021-125627OB-C32, from the Centre of Excellence “Severo Ochoa'' award to the Instituto de Astrofísica de Canarias (CEX2019-000920-S), from the Centre of Excellence “María de Maeztu” award to the Institut de Ciències de l’Espai (CEX2020-001058-M), and from the Generalitat de Catalunya/CERCA programme. 
SCCB acknowledges the support from Fundação para a Ciência e Tecnologia (FCT) in the form of work contract through the Scientific Employment Incentive program with reference 2023.06687.CEECIND. 
ACC acknowledges support from STFC consolidated grant number ST/V000861/1, and UKSA grant number ST/X002217/1. 
ACMC acknowledges support from the FCT, Portugal, through the CFisUC projects UIDB/04564/2020 and UIDP/04564/2020, with DOI identifiers 10.54499/UIDB/04564/2020 and 10.54499/UIDP/04564/2020, respectively. 
A.C., A.D., B.E., K.G., and J.K. acknowledge their role as ESA-appointed CHEOPS Science Team Members. 
P.E.C. is funded by the Austrian Science Fund (FWF) Erwin Schroedinger Fellowship, program J4595-N. 
This project was supported by the CNES. 
This work was supported by FCT - Funda\c{c}\~{a}o para a Ci\^{e}ncia e a Tecnologia through national funds and by FEDER through COMPETE2020 through the research grants UIDB/04434/2020, UIDP/04434/2020, 2022.06962.PTDC. 
O.D.S.D. is supported in the form of work contract (DL 57/2016/CP1364/CT0004) funded by national funds through FCT. 
B.-O. D. acknowledges support from the Swiss State Secretariat for Education, Research and Innovation (SERI) under contract number MB22.00046. 
ADe, BEd, KGa, and JKo acknowledge their role as ESA-appointed CHEOPS Science Team Members. 
C.B. acknowledges support from the Swiss NCCR PlanetS. 
This work has been carried out within the framework of the NCCR PlanetS supported by the Swiss National Science Foundation under grants 51NF40182901 and 51NF40205606. J.K. acknowledges support from the Swiss National Science Foundation under grant number TMSGI2\_211697.
MF and CMP gratefully acknowledge the support of the Swedish National Space Agency (DNR 65/19, 174/18). 
DG gratefully acknowledges financial support from the CRT foundation under Grant No. 2018.2323 “Gaseousor rocky? Unveiling the nature of small worlds”. 
M.G. is an F.R.S.-FNRS Senior Research Associate. 
CHe acknowledges the European Union H2020-MSCA-ITN-2019 under GrantAgreement no. 860470 (CHAMELEON), and the HPC facilities at the Vienna Science Cluster (VSC). 
MNG is the ESA CHEOPS Project Scientist and Mission Representative. BMM is the ESA CHEOPS Project Scientist. KGI was the ESA CHEOPS Project Scientist until the end of December 2022 and Mission Representative until the end of January 2023. All of them are/were responsible for the Guest Observers (GO) Programme. None of them relay/relayed proprietary information between the GO and Guaranteed Time Observation (GTO) Programmes, nor do/did they decide on the definition and target selection of the GTO Programme.
K.W.F.L. was supported by Deutsche Forschungsgemeinschaft grants RA714/14-1 within the DFG Schwerpunkt SPP 1992, Exploring the Diversity of Extrasolar Planets. 
This work was granted access to the HPC resources of MesoPSL financed by the Region Ile de France and the project Equip@Meso (reference ANR-10-EQPX-29-01) of the programme Investissements d'Avenir supervised by the Agence Nationale pour la Recherche. 
This work has been carried out within the framework of the NCCR PlanetS supported by the Swiss National Science Foundation under grants 51NF40\_182901 and 51NF40\_205606. AL acknowledges support of the Swiss National Science Foundation under grant number  TMSGI2\_211697. 
ML acknowledges support of the Swiss National Science Foundation under grant number PCEFP2\_194576. 
PM acknowledges support from STFC research grant number ST/R000638/1. 
This work was also partially supported by a grant from the Simons Foundation (PI Queloz, grant number 327127). 
NCSa acknowledges funding by the European Union (ERC, FIERCE, 101052347). Views and opinions expressed are however those of the author(s) only and do not necessarily reflect those of the European Union or the European Research Council. Neither the European Union nor the granting authority can be held responsible for them. 
V.V.G. is an F.R.S-FNRS Research Associate. 
JV acknowledges support from the Swiss National Science Foundation (SNSF) under grant PZ00P2\_208945. 
EV acknowledges support from the ‘DISCOBOLO’ project funded by the Spanish Ministerio de Ciencia, Innovación y Universidades undergrant PID2021-127289NB-I00. 
NAW acknowledges UKSA grant ST/R004838/1. 

Project no. C1746651 has been implemented with the support provided by the Ministry of Culture and Innovation of Hungary from the National Research, Development and Innovation Fund, financed under the NVKDP-2021 funding scheme.

\end{acknowledgements}

\bibliography{refs}

\begin{appendix}
\onecolumn

\section{Information about the CHEOPS observations}
The metadata of the four CHEOPS observations are shown in Table \ref{tab:visits}. The list of nearby stars, that may interfere with the aperture photometry, is shown in Table \ref{tab:starcat}.
\begin{table}[!h]
    \caption{File keys and the respective observation dates for the four CHEOPS visits of HD 95338.}   \label{tab:visits}
    \centering
    \begin{tabular}{c c c c c}
    \hline
        File key & Observation start & Observation end & Number of orbits & Efficiency [\%]\\
        \hline
        \hline
        \verb|CH_PR100009_TG001101| & 2021-03-26T23:27:04 & 2021-03-28T17:43:29 & 26 & 59.5\\
       \verb|CH_PR100009_TG001102|  & 2022-04-16T13:53:05 & 2022-04-18T08:37:02 & 26 & 57.6\\ 
       \verb|CH_PR140072_TG000901| & 2023-05-07T03:44:19 & 2023-05-09T01:44:46 & 28 & 58.2\\
       \verb|CH_PR140072_TG000902| & 2025-04-22T06:55:00 & 2025-04-24T04:53:43 & 29 & 64.1 \\
       \hline
    \end{tabular}
\end{table}

\begin{table}[!h]
    \centering   
    \caption{List of stars that are ($i$) within $100"$ of HD 95338 and ($ii$) brighter than $G = 15.2$ magnitudes.} \label{tab:starcat}
    \begin{tabular}{l c c c c}
    \hline
    Gaia ID & $\alpha_{J2000}$ [$^\circ$] &  $\delta_{J2000}$ [$^\circ$] & Separation from target [$"$] & $G$ [mag] \\
    \hline
    \hline
    Gaia DR2 5340648488081462528\tablefootmark{a}   &  164.854663528690 & -56.6238174607725 &0 & 8.38 \\
    Gaia DR2 5340648110124332288   &  164.849734257485 &-56.6295150776728 &25.1166801452637 & 14.19 \\
    Gaia DR2 5340648110124331008   &  164.863838643089 &-56.6369880728433 &50.5965957641602 & 14.66 \\
    Gaia DR2 5340648110124328448   &  164.864406506541 &-56.6401925897662 &61.9967308044434 & 14.14 \\
    Gaia DR2 5340648621192205696   &  164.876026962042 &-56.6081809926106 &67.7202606201172 & 14.80 \\
    Gaia DR2 5340648075764581888   &  164.847011556551 &-56.6425190710745 &70.7192840576172 & 14.45 \\
    Gaia DR2 5340647766526968704   &  164.892506952765 &-56.6236315284767 &71.6561660766602 & 12.77 \\
    Gaia DR2 5340648110124325248   &  164.862354443584 &-56.643234452212 &71.8457717895508 & 13.71 \\
    Gaia DR2 5340648591142853888   &  164.851492797189 &-56.5991502768792 &88.3759384155273 & 10.57 \\
    Gaia DR2 5340648140155761536   &  164.820502759141 &-56.639273910009 &90.7409591674805 & 14.95 \\
    Gaia DR2 5340647354210316288   &  164.874514445752 &-56.6468253286776 & 91.1862640380859 & 15.00 \\
    Gaia DR2 5340648591160693376   &  164.842872932682 &-56.5989572030799 &92.4748764038086 & 15.10 \\
    Gaia DR2 5340648762959367168   &  164.815444234184 &-56.6093308787378 &95.8126449584961 &14.09 \\
    \hline
     \end{tabular}
   \tablefoot{\tablefoottext{a}{Target star, HD 95338.}}
\end{table}

    \section{Sensitivity maps}
    \subsection{Quadratic trend}
\begin{figure}[!h]
        \centering
        \includegraphics[width = \textwidth]{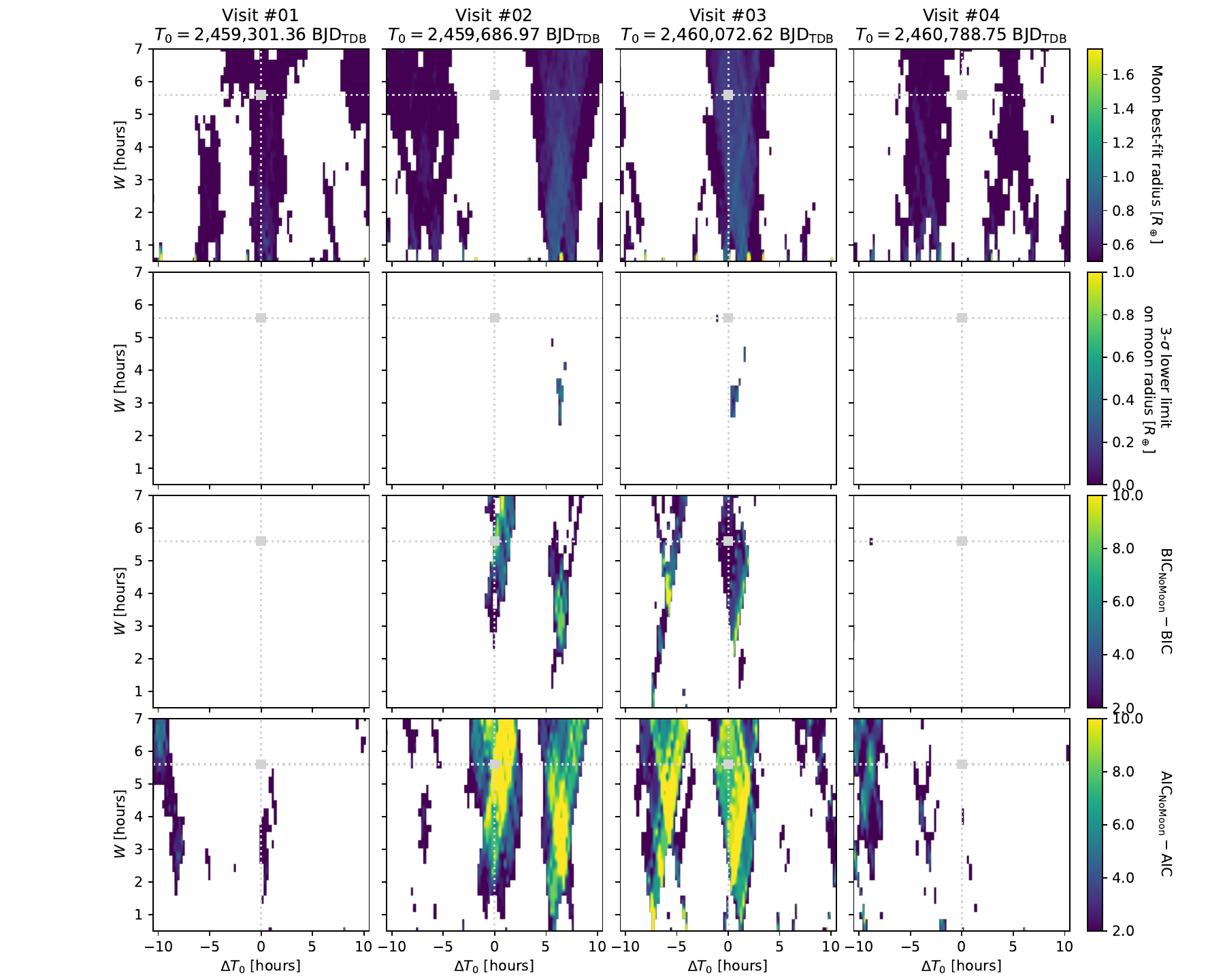}
        \caption{Same as Fig. \ref{fig:sensitivity-maps} but with a quadratic trend instead of a linear trend.}
        \label{fig:drp25_def0_senmap_quad}
    \end{figure}

    The sensitivity maps in case of a quadratic trend (for the real observations) are shown on Fig. \ref{fig:drp25_def0_senmap_quad}.
    \clearpage
    \subsection{Synthetic transits}
\begin{figure}[!h]
        \centering
        \includegraphics[width = \textwidth]{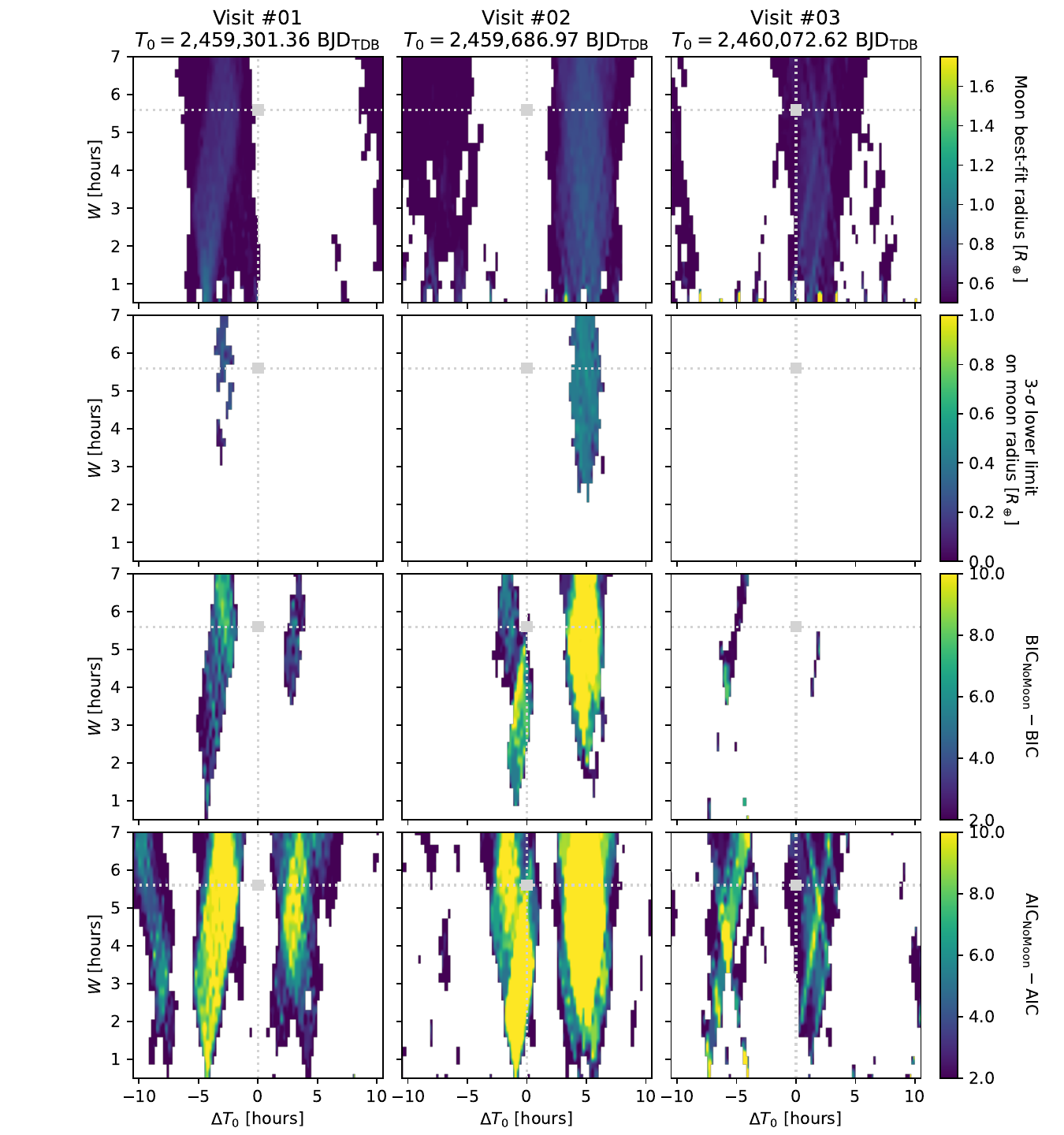}
        \caption{Same as Fig. \ref{fig:sensitivity-maps} but with a quadratic trend instead of a linear trend and with an $R_{\rm Moon} = 0.8$~R$\oplus$ moon injected into the first two visits. The injection-and retrieval tests were carried out only on the first three CHEOPS visits.}
        \label{fig:drp25_def16_senmap_quad}
    \end{figure}

\begin{figure}[!h]
        \centering
      
        \includegraphics[width = \textwidth]{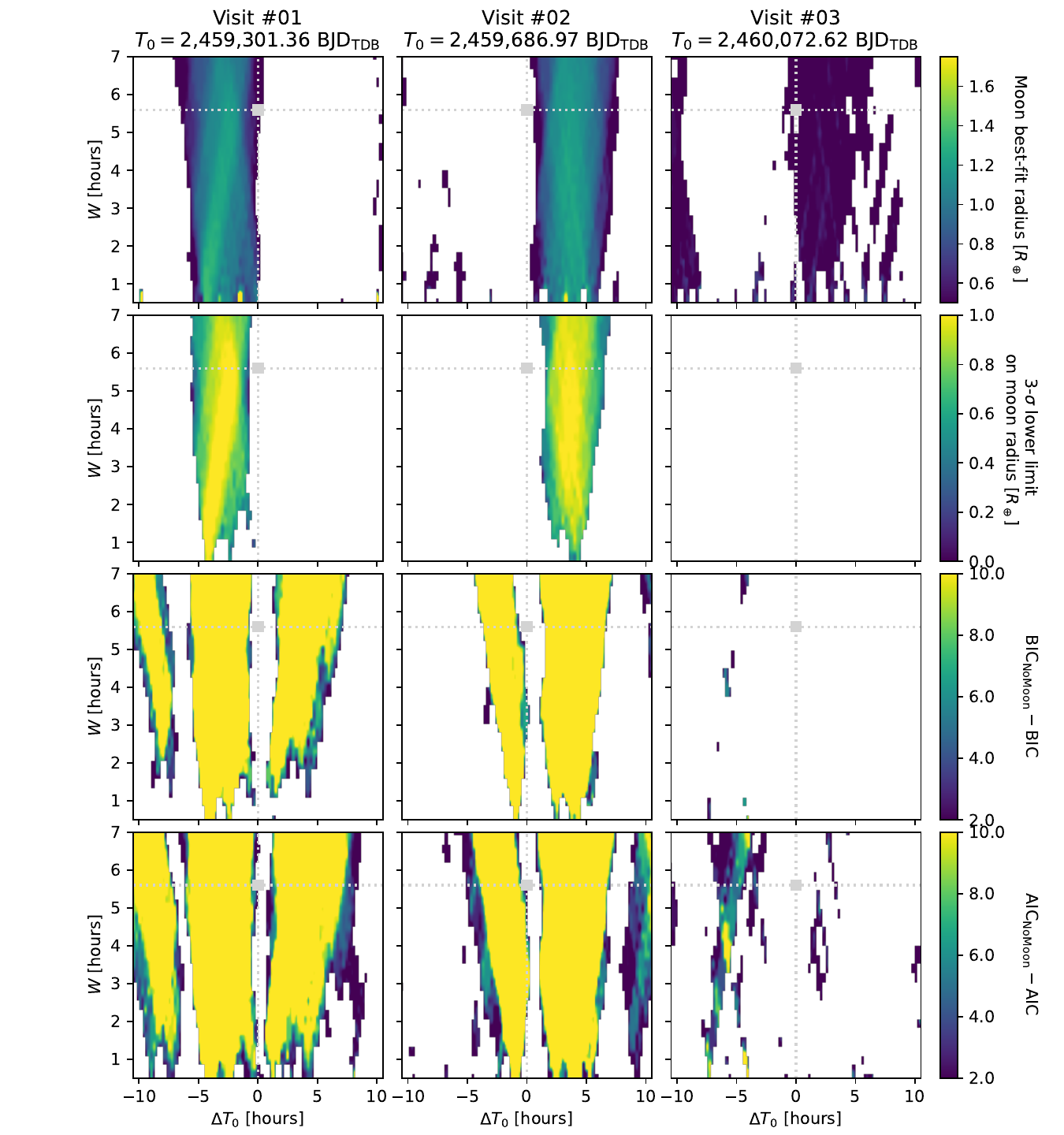}
        \caption{Same as Fig. \ref{fig:drp25_def16_senmap_quad} but with a quadratic trend instead of a linear trend and with an $R_{\rm Moon} = 1.35$~R$\oplus$ moon injected into the first two visits.}
        \label{fig:drp25_def27_senmap_quad}
    \end{figure}

\begin{figure}[!h]
        \centering
        \includegraphics[width = \textwidth]{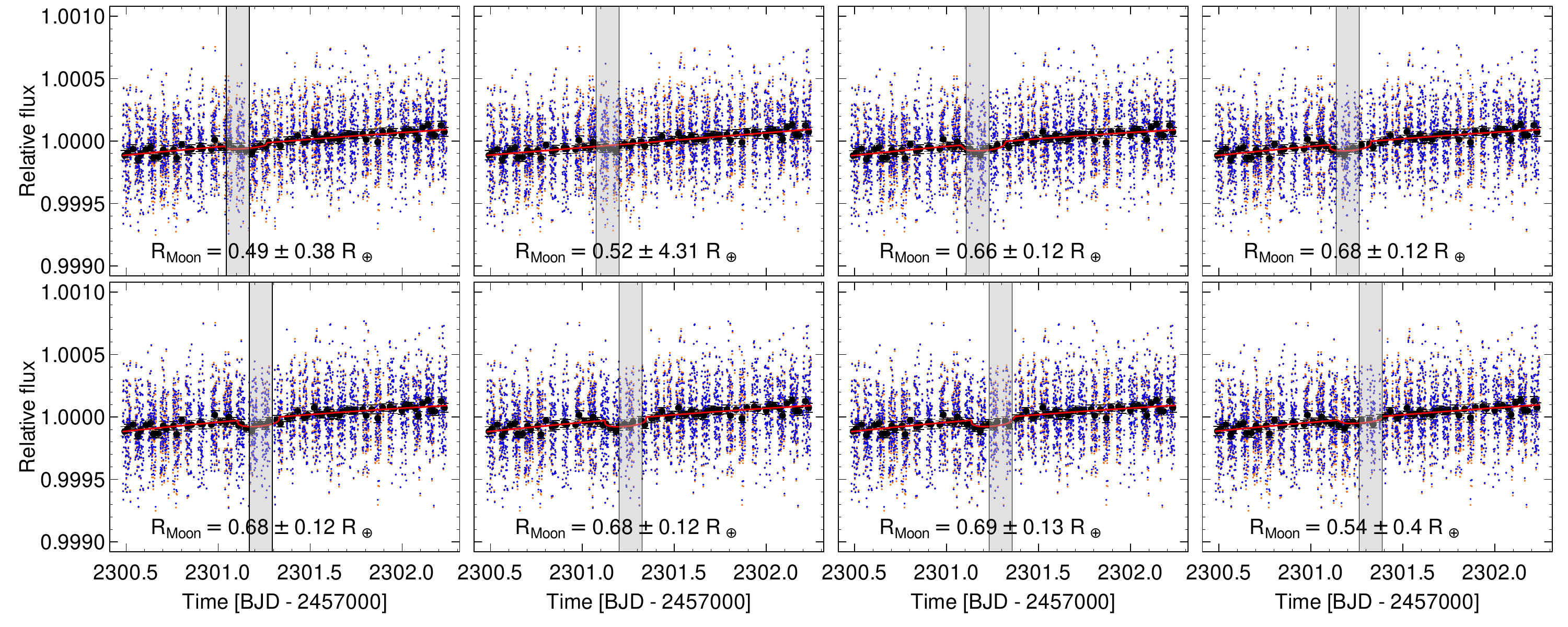} \\
        \includegraphics[width = \textwidth]{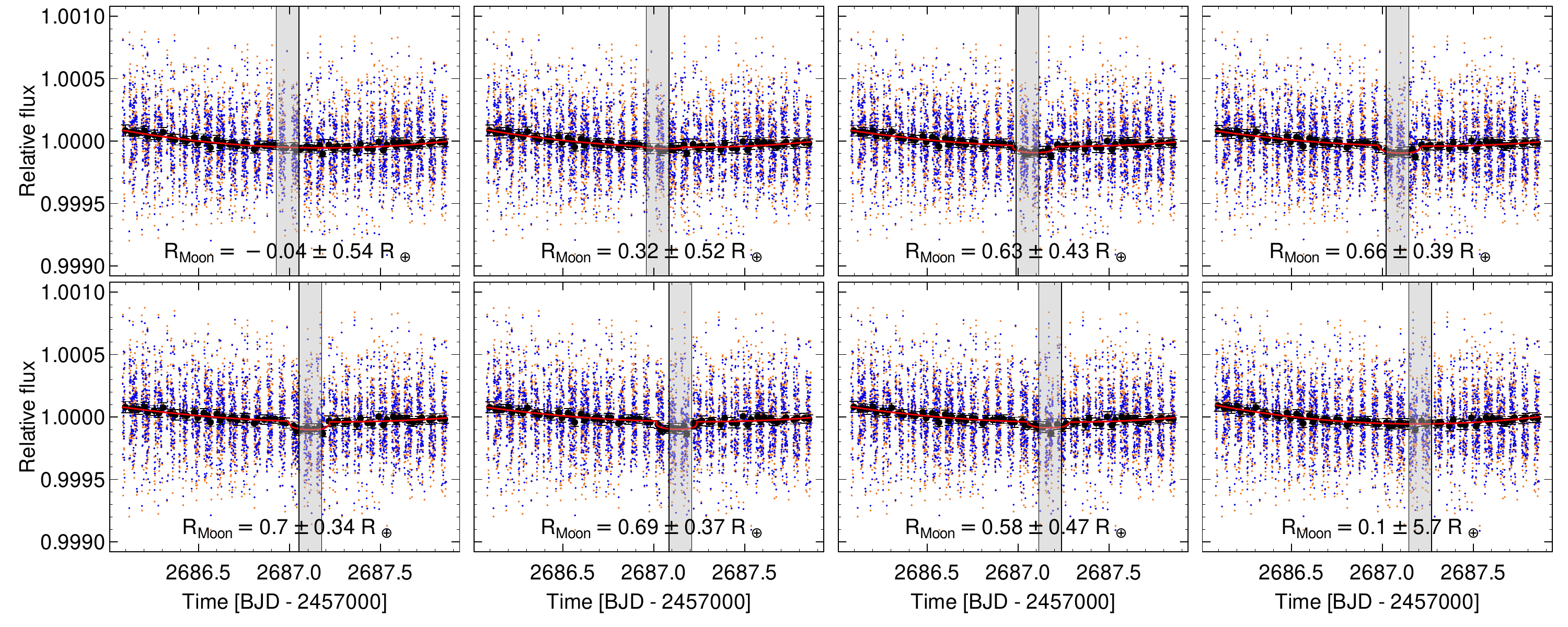}
      
        \caption{Performance of the box-shifting approach on the injected exomoon transits. \textbf{Top rows:} Recovery of an $R_{\rm Moon} = 0.8$~R$_\oplus$ in the first visit. \textbf{Bottom row:} Recovery of an $R_{\rm Moon} = 0.8$~R$_\oplus$ in the second visit.}
        \label{fig:boxshift_injections_def16}
    \end{figure}

\begin{figure}[!h]
        \centering
        \includegraphics[width = \textwidth]{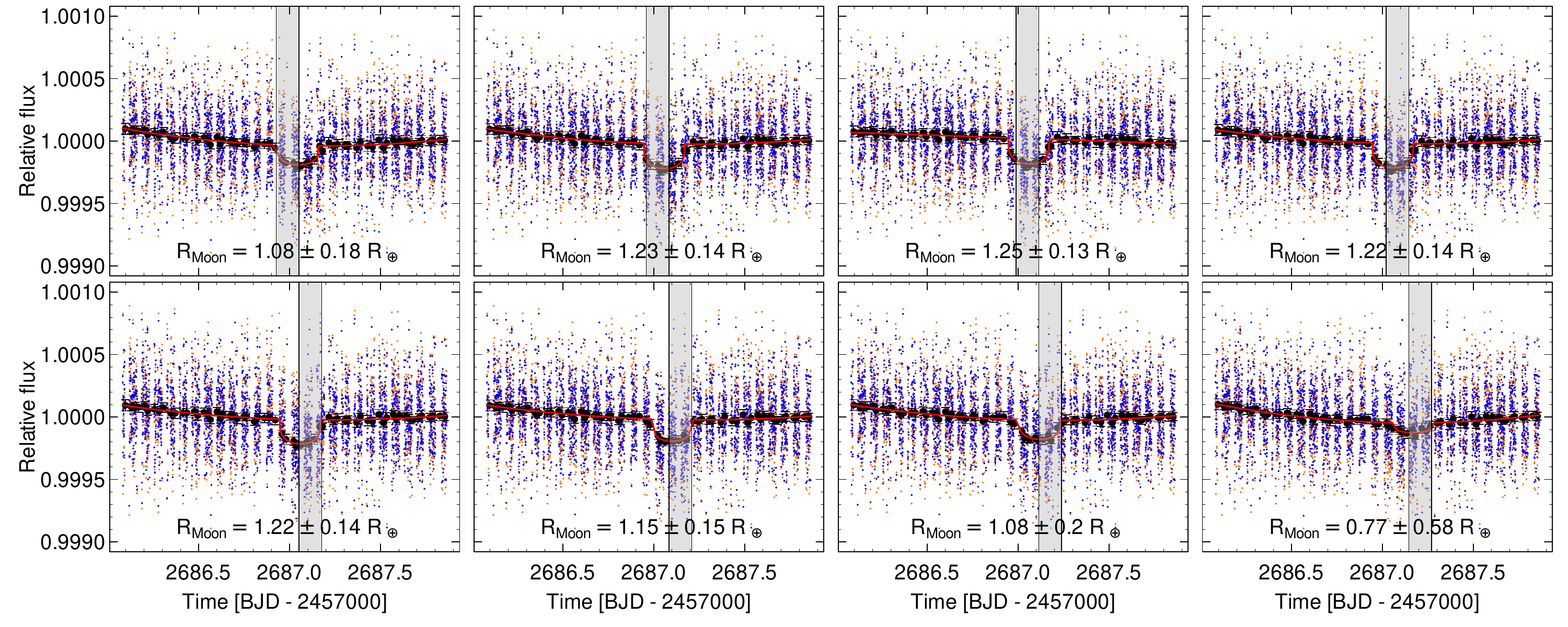} \\
        \includegraphics[width = \textwidth]{def27_visit2_quad_new.pdf} \\
        \caption{Performance of the box-shifting approach on the injected exomoon transits. \textbf{Top row:} Recovery of an $R_{\rm Moon} = 1.35$~R$_\oplus$ in the first visit. \textbf{Bottom row:} Recovery of an $R_{\rm Moon} = 1.35$~R$_\oplus$ in the second visit.}
        \label{fig:boxshift_injections_def27}
    \end{figure}

    The results of injection-and-retrieval tests are shown in Figs. \ref{fig:drp25_def16_senmap_quad} -- \ref{fig:boxshift_injections_def27}, and Tables \ref{tab:recovery_visit1} -- \ref{tab:recovery_visit1}.

\begin{table}[!h]
\caption{Times of midtransit and recovered radii by the box-shifting method, of the injection-and-retrieval tests ($T_{0, \textrm{Moon, injected}} = 2301.25$) from the first visit.}
\label{tab:recovery_visit1}
\centering
\scriptsize
\begin{tabular}{l c c c c c c}
\hline
\hline
Searchbox number & Searchbox left edge & Searchbox right edge & $T_{0, \textrm{Moon}}$ & $R_{\rm Moon}$~[$R_\oplus$] & $\frac{T_{0, \textrm{Moon, injected}}-T_{0, \textrm{Moon}}}{\Delta T_{0, \textrm{Moon}}}$ & $\frac{R_{\rm Moon, injected} - R_{\rm Moon}}{\Delta R_{\rm Moon} }$  \\
\hline
\multicolumn{7}{c}{$R_{\rm Moon, injected} = 0.8$~$R_\oplus$} \\
\hline
1 &  $2301.0438$ & $2301.1688$ & $2301.152 \pm 0.039$ & $0.47 \pm 0.36$ & $2.5$ & $0.9$ \\ 
2 &  $2301.0750$ & $2301.2000$ & $2301.179 \pm 0.019$ & $0.60 \pm 0.15$ & $3.8$ & $1.3$ \\ 
3 &  $2301.1063$ & $2301.2313$ & $2301.202 \pm 0.023$ & $0.63 \pm 0.13$ & $2.1$ & $1.4$ \\ 
4 &  $2301.1375$ & $2301.2625$ & $2301.216 \pm 0.032$ & $0.64 \pm 0.12$ & $1.1$ & $1.2$ \\ 
5 &  $2301.1688$ & $2301.2938$ & $2301.225 \pm 0.030$ & $0.65 \pm 0.12$ & $0.8$ & $1.3$ \\ 
6 &  $2301.2000$ & $2301.3250$ & $2301.238 \pm 0.024$ & $0.66 \pm 0.12$ & $0.5$ & $1.2$ \\ 
7 &  $2301.2313$ & $2301.3563$ & $2301.246 \pm 0.014$ & $0.66 \pm 0.13$ & $0.3$ & $1.0$ \\ 
8 &  $2301.2625$ & $2301.3875$ & $2301.276 \pm 0.038$ & $0.52 \pm 0.42$ & $-0.7$ & $0.7$ \\ 
\hline
\multicolumn{7}{c}{$R_{\rm Moon, injected} = 0.9$~$R_\oplus$} \\
\hline
1 &  $2301.0438$ & $2301.1688$ & $2301.154 \pm 0.034$ & $0.55 \pm 0.34$ & $2.8$ & $1.0$ \\ 
2 &  $2301.0750$ & $2301.2000$ & $2301.181 \pm 0.017$ & $0.66 \pm 0.13$ & $4.2$ & $1.9$ \\ 
3 &  $2301.1063$ & $2301.2313$ & $2301.209 \pm 0.022$ & $0.71 \pm 0.11$ & $1.9$ & $1.7$ \\ 
4 &  $2301.1375$ & $2301.2625$ & $2301.238 \pm 0.021$ & $0.75 \pm 0.11$ & $0.6$ & $1.3$ \\ 
5 &  $2301.1688$ & $2301.2938$ & $2301.239 \pm 0.022$ & $0.75 \pm 0.11$ & $0.5$ & $1.4$ \\ 
6 &  $2301.2000$ & $2301.3250$ & $2301.240 \pm 0.019$ & $0.75 \pm 0.11$ & $0.5$ & $1.3$ \\ 
7 &  $2301.2313$ & $2301.3563$ & $2301.243 \pm 0.010$ & $0.77 \pm 0.10$ & $0.7$ & $1.2$ \\ 
8 &  $2301.2625$ & $2301.3875$ & $2301.269 \pm 0.008$ & $0.71 \pm 0.16$ & $-2.4$ & $1.2$ \\ 
\hline
\multicolumn{7}{c}{$R_{\rm Moon, injected} = 1.05$~$R_\oplus$} \\
\hline
1 &  $2301.0438$ & $2301.1688$ & $2301.158 \pm 0.019$ & $0.68 \pm 0.22$ & $4.8$ & $1.7$ \\ 
2 &  $2301.0750$ & $2301.2000$ & $2301.183 \pm 0.014$ & $0.77 \pm 0.12$ & $4.8$ & $2.4$ \\ 
3 &  $2301.1063$ & $2301.2313$ & $2301.222 \pm 0.013$ & $0.85 \pm 0.10$ & $2.2$ & $1.9$ \\ 
4 &  $2301.1375$ & $2301.2625$ & $2301.241 \pm 0.006$ & $0.93 \pm 0.09$ & $1.6$ & $1.4$ \\ 
5 &  $2301.1688$ & $2301.2938$ & $2301.241 \pm 0.006$ & $0.93 \pm 0.09$ & $1.5$ & $1.4$ \\ 
6 &  $2301.2000$ & $2301.3250$ & $2301.241 \pm 0.006$ & $0.93 \pm 0.09$ & $1.5$ & $1.3$ \\ 
7 &  $2301.2313$ & $2301.3563$ & $2301.242 \pm 0.006$ & $0.93 \pm 0.09$ & $1.4$ & $1.3$ \\ 
8 &  $2301.2625$ & $2301.3875$ & $2301.267 \pm 0.004$ & $0.88 \pm 0.10$ & $-3.8$ & $1.6$ \\ 
\hline
\multicolumn{7}{c}{$R_{\rm Moon, injected} = 1.20$~$R_\oplus$} \\
\hline
1 &  $2301.0438$ & $2301.1688$ & $2301.160 \pm 0.010$ & $0.77 \pm 0.18$ & $9.0$ & $2.4$ \\ 
2 &  $2301.0750$ & $2301.2000$ & $2301.189 \pm 0.011$ & $0.87 \pm 0.11$ & $5.5$ & $3.1$ \\ 
3 &  $2301.1063$ & $2301.2313$ & $2301.229 \pm 0.004$ & $1.02 \pm 0.09$ & $5.4$ & $2.1$ \\ 
4 &  $2301.1375$ & $2301.2625$ & $2301.241 \pm 0.003$ & $1.09 \pm 0.08$ & $2.8$ & $1.4$ \\ 
5 &  $2301.1688$ & $2301.2938$ & $2301.241 \pm 0.003$ & $1.09 \pm 0.08$ & $2.8$ & $1.4$ \\ 
6 &  $2301.2000$ & $2301.3250$ & $2301.241 \pm 0.003$ & $1.09 \pm 0.08$ & $2.8$ & $1.4$ \\ 
7 &  $2301.2313$ & $2301.3563$ & $2301.241 \pm 0.003$ & $1.09 \pm 0.07$ & $2.8$ & $1.5$ \\ 
8 &  $2301.2625$ & $2301.3875$ & $2301.265 \pm 0.003$ & $1.04 \pm 0.09$ & $-5.1$ & $1.8$ \\
\hline
\multicolumn{7}{c}{$R_{\rm Moon, injected} = 1.35$~$R_\oplus$} \\
\hline
1 &  $2301.0438$ & $2301.1688$ & $2301.163 \pm 0.007$ & $0.87 \pm 0.14$ & $12.9$ & $3.5$ \\ 
2 &  $2301.0750$ & $2301.2000$ & $2301.195 \pm 0.010$ & $0.97 \pm 0.10$ & $5.5$ & $3.7$ \\ 
3 &  $2301.1063$ & $2301.2313$ & $2301.230 \pm 0.002$ & $1.15 \pm 0.08$ & $10.9$ & $2.5$ \\ 
4 &  $2301.1375$ & $2301.2625$ & $2301.241 \pm 0.002$ & $1.23 \pm 0.07$ & $3.9$ & $1.8$ \\ 
5 &  $2301.1688$ & $2301.2938$ & $2301.241 \pm 0.002$ & $1.23 \pm 0.07$ & $3.8$ & $1.8$ \\ 
6 &  $2301.2000$ & $2301.3250$ & $2301.241 \pm 0.002$ & $1.23 \pm 0.07$ & $3.8$ & $1.7$ \\ 
7 &  $2301.2313$ & $2301.3563$ & $2301.241 \pm 0.002$ & $1.23 \pm 0.07$ & $3.8$ & $1.7$ \\ 
8 &  $2301.2625$ & $2301.3875$ & $2301.264 \pm 0.002$ & $1.17 \pm 0.08$ & $-6.7$ & $2.1$ \\ 
\hline
\end{tabular}
\end{table}

\begin{table}[!h]
\caption{Times of midtransit and recovered radii by the box-shifting method, of the injection-and-retrieval tests ($T_{0, \textrm{Moon, injected}} = 2687.05$ from the first visit. }
\label{tab:recovery_visit2}
\centering
\scriptsize
\begin{tabular}{l c c c c c c}
\hline
\hline
Searchbox number & Searchbox left edge & Searchbox right edge & $T_{0, \textrm{Moon}}$ & $R_{\rm Moon}$~[$R_\oplus]$ & $\frac{T_{0, \textrm{Moon, injected}}-T_{0, \textrm{Moon}}}{\Delta T_{0, \textrm{Moon}}}$ & $\frac{R_{\rm Moon, injected} - R_{\rm Moon}}{\Delta R_{\rm Moon} }$  \\
\hline
\multicolumn{7}{c}{$R_{\rm Moon, injected} = 0.8$~$R_\oplus$} \\
\hline
1 &  $2686.9265$ & $2687.0515$ & $2686.962 \pm 0.048$ & $-0.28 \pm 0.55$ & $1.8$ & $2.0$ \\ 
2 &  $2686.9578$ & $2687.0828$ & $2687.068 \pm 0.033$ & $0.60 \pm 0.39$ & $-0.5$ & $0.5$ \\ 
3 &  $2686.9890$ & $2687.1140$ & $2687.089 \pm 0.020$ & $0.70 \pm 0.20$ & $-2.0$ & $0.5$ \\ 
4 &  $2687.0203$ & $2687.1453$ & $2687.100 \pm 0.024$ & $0.70 \pm 0.18$ & $-2.0$ & $0.5$ \\ 
5 &  $2687.0515$ & $2687.1765$ & $2687.106 \pm 0.036$ & $0.72 \pm 0.17$ & $-1.6$ & $0.4$ \\ 
6 &  $2687.0828$ & $2687.2078$ & $2687.118 \pm 0.034$ & $0.71 \pm 0.18$ & $-2.0$ & $0.5$ \\ 
7 &  $2687.1140$ & $2687.2390$ & $2687.152 \pm 0.028$ & $0.69 \pm 0.22$ & $-3.6$ & $0.5$ \\ 
8 &  $2687.1453$ & $2687.2703$ & $2687.170 \pm 0.028$ & $0.62 \pm 0.35$ & $-4.3$ & $0.5$ \\ 
\hline
\multicolumn{7}{c}{$R_{\rm Moon, injected} = 0.9$~$R_\oplus$} \\
\hline
1 &  $2686.9265$ & $2687.0515$ & $2687.002 \pm 0.054$ & $0.14 \pm 0.64$ & $0.9$ & $1.2$ \\ 
2 &  $2686.9578$ & $2687.0828$ & $2687.075 \pm 0.013$ & $0.77 \pm 0.18$ & $-1.8$ & $0.7$ \\ 
3 &  $2686.9890$ & $2687.1140$ & $2687.090 \pm 0.015$ & $0.80 \pm 0.14$ & $-2.7$ & $0.7$ \\ 
4 &  $2687.0203$ & $2687.1453$ & $2687.096 \pm 0.019$ & $0.80 \pm 0.14$ & $-2.4$ & $0.7$ \\ 
5 &  $2687.0515$ & $2687.1765$ & $2687.099 \pm 0.024$ & $0.80 \pm 0.14$ & $-2.0$ & $0.7$ \\ 
6 &  $2687.0828$ & $2687.2078$ & $2687.104 \pm 0.026$ & $0.80 \pm 0.15$ & $-2.1$ & $0.7$ \\ 
7 &  $2687.1140$ & $2687.2390$ & $2687.141 \pm 0.026$ & $0.75 \pm 0.18$ & $-3.5$ & $0.8$ \\ 
8 &  $2687.1453$ & $2687.2703$ & $2687.168 \pm 0.027$ & $0.67 \pm 0.37$ & $-4.4$ & $0.6$ \\ 
\hline
\multicolumn{7}{c}{$R_{\rm Moon, injected} = 1.05$~$R_\oplus$} \\
\hline
1 &  $2686.9265$ & $2687.0515$ & $2687.038 \pm 0.024$ & $0.72 \pm 0.36$ & $0.5$ & $0.9$ \\ 
2 &  $2686.9578$ & $2687.0828$ & $2687.075 \pm 0.012$ & $0.91 \pm 0.13$ & $-2.1$ & $1.1$ \\ 
3 &  $2686.9890$ & $2687.1140$ & $2687.084 \pm 0.017$ & $0.90 \pm 0.12$ & $-2.0$ & $1.2$ \\ 
4 &  $2687.0203$ & $2687.1453$ & $2687.085 \pm 0.017$ & $0.90 \pm 0.12$ & $-2.1$ & $1.2$ \\ 
5 &  $2687.0515$ & $2687.1765$ & $2687.086 \pm 0.017$ & $0.90 \pm 0.12$ & $-2.1$ & $1.2$ \\ 
6 &  $2687.0828$ & $2687.2078$ & $2687.101 \pm 0.015$ & $0.90 \pm 0.13$ & $-3.5$ & $1.2$ \\ 
7 &  $2687.1140$ & $2687.2390$ & $2687.125 \pm 0.020$ & $0.86 \pm 0.16$ & $-3.7$ & $1.2$ \\ 
8 &  $2687.1453$ & $2687.2703$ & $2687.164 \pm 0.019$ & $0.74 \pm 0.29$ & $-6.0$ & $1.0$ \\ 
\hline
\multicolumn{7}{c}{$R_{\rm Moon, injected} = 1.20$~$R_\oplus$} \\
\hline
1 &  $2686.9265$ & $2687.0515$ & $2687.048 \pm 0.010$ & $0.95 \pm 0.13$ & $0.2$ & $1.8$ \\ 
2 &  $2686.9578$ & $2687.0828$ & $2687.069 \pm 0.012$ & $1.05 \pm 0.11$ & $-1.7$ & $1.3$ \\ 
3 &  $2686.9890$ & $2687.1140$ & $2687.076 \pm 0.018$ & $1.05 \pm 0.11$ & $-1.5$ & $1.4$ \\ 
4 &  $2687.0203$ & $2687.1453$ & $2687.075 \pm 0.020$ & $1.04 \pm 0.11$ & $-1.3$ & $1.5$ \\ 
5 &  $2687.0515$ & $2687.1765$ & $2687.077 \pm 0.020$ & $1.04 \pm 0.11$ & $-1.3$ & $1.4$ \\ 
6 &  $2687.0828$ & $2687.2078$ & $2687.098 \pm 0.010$ & $1.02 \pm 0.11$ & $-4.7$ & $1.6$ \\ 
7 &  $2687.1140$ & $2687.2390$ & $2687.119 \pm 0.010$ & $0.97 \pm 0.14$ & $-6.9$ & $1.7$ \\ 
8 &  $2687.1453$ & $2687.2703$ & $2687.162 \pm 0.021$ & $0.79 \pm 0.36$ & $-5.3$ & $1.1$ \\ 
\hline
\multicolumn{7}{c}{$R_{\rm Moon, injected} = 1.35$~$R_\oplus$} \\
\hline
1 &  $2686.9265$ & $2687.0515$ & $2687.049 \pm 0.002$ & $1.14 \pm 0.11$ & $0.3$ & $1.9$ \\ 
2 &  $2686.9578$ & $2687.0828$ & $2687.057 \pm 0.006$ & $1.20 \pm 0.10$ & $-1.2$ & $1.5$ \\ 
3 &  $2686.9890$ & $2687.1140$ & $2687.057 \pm 0.010$ & $1.20 \pm 0.10$ & $-0.7$ & $1.5$ \\ 
4 &  $2687.0203$ & $2687.1453$ & $2687.057 \pm 0.009$ & $1.20 \pm 0.10$ & $-0.8$ & $1.5$ \\ 
5 &  $2687.0515$ & $2687.1765$ & $2687.057 \pm 0.009$ & $1.20 \pm 0.10$ & $-0.8$ & $1.5$ \\ 
6 &  $2687.0828$ & $2687.2078$ & $2687.092 \pm 0.009$ & $1.13 \pm 0.11$ & $-4.5$ & $1.9$ \\ 
7 &  $2687.1140$ & $2687.2390$ & $2687.117 \pm 0.004$ & $1.09 \pm 0.13$ & $-19.0$ & $2.0$ \\ 
8 &  $2687.1453$ & $2687.2703$ & $2687.161 \pm 0.023$ & $0.82 \pm 0.41$ & $-4.9$ & $1.3$ \\ 
\hline
\end{tabular}
\end{table}

\end{appendix}

\end{document}